\documentclass[showpacs]{interact}

\usepackage{epstopdf}
\usepackage{subfigure}

\usepackage[numbers,sort&compress,merge]{natbib}

\theoremstyle{plain}

\theoremstyle{definition}

\theoremstyle{remark}

\usepackage{braket}
\usepackage{color,soul}
\begin{document}

\title{Effects of partial measurements on quantum resources and quantum Fisher information of a teleported state in a relativistic scenario }

\author{
	\name{M. Jafarzadeh\textsuperscript{a}, H. Rangani Jahromi\textsuperscript{b}\thanks{Corresponding author: H. Rangani Jahromi. Email: h.ranganijahromi@jahromu.ac.ir}, and M. Amniat-Talab\textsuperscript{a}}
	\affil{\textsuperscript{a}Physics Department, Faculty of Sciences, Urmia University, P.B. 165, Urmia, Iran; \textsuperscript
		{b}Physics Department, Faculty of Sciences, Jahrom University, P.B. 74135111, Jahrom, Iran.}
}

\maketitle
\begin{abstract}
We address the teleportation of single- and two-qubit quantum states, parametrized by weight $\theta$ and phase $\phi$ parameters, in the presence of the Unruh effect experienced by a mode of a free Dirac field. We investigate the effects of the partial measurement (PM) and partial measurement reversal (PMR) 
on the quantum resources (QRs) and quantum Fisher information (QFI) of the teleported states. In particular,
we discuss  the optimal behavior of the QFI, quantum coherence (QC) as well as fidelity with respect to the PM and PMR strength and examine
the effect of the Unruh noise on optimal estimation. It is found that in the single-qubit scenario, the PM (PMR) strength at which the optimal 
estimation of the phase parameter occurs, is the same as the PM (PMR) strength with which the teleportation fidelity and the QC of the teleported 
single-qubit state reaches to its maximum value. On the other hand, generalizing the results to two-qubit teleportation, we find that the encoded
information in the weight parameter is better protected against the Unruh noise in two-qubit teleportation than the one-qubit scenario. 
However, extraction of information encoded in the phase parameter is more efficient in single-qubit teleportation than the two-qubit one.
\end{abstract}


\maketitle
\section{Introduction}

Quantum teleportation~\cite{C. H. Bennett} is undoubtedly one of the most striking implications predicted by quantum mechanics and it is an important ingredient for quantum communication and quantum information processing (QIP) \cite{D. Gottesman,R. Ursin}. In the last decades theoretical and experimental consideration of quantum teleportation has attracted many researchers' attention~\cite{S.L. Braunstein,D. Bouwmeester,A. Furusawa,Y. H. Kim,R. Riebe,F. DellAnno,S. Adhikari,Sk. Sazim,X. L. Wang,Jafarzadeh M., STM+13, HDC+13,JPH+17,RG18,KJC+18}. Quantum teleportation is described as a process by which an arbitrary unknown quantum state can be transmitted faithfully from one object to another, without physical traveling of the object itself. The system is isolated from the external forces in the original form of the teleportation \cite{C. H. Bennett}, and a maximally entangled pair is used as the resource. However, decoherence \cite{W.H. Zurek,M. Schlosshawer} is an inevitable phenomenon in open quantum systems which takes place due to the interaction between the system and environment. This leads to the degradation of quantum correlations, a fundamental resource for QIP, and therefore influences the fidelity in quantum state teleportation \cite{S. Oh, E. Jung, D. D. Bhaktavatsala Rao}.

Relativistic quantum information (RQI) \cite{P. Alsing,R. B. Mann} aims to realize the relationship between relativity as well as quantum information, and combine relativistic effects to amend quantum information tasks, e.g., quantum teleportation. Moreover, in RQI we try to understand how these protocols may be realized in curved space time. Unruh effect \cite{Unruh WG,Birrell ND}, a significant prediction in quantum field theory, proposes that a uniformly accelerated observer in Minkowski spacetime (Rindler observer) associates a thermal bath of Rindler particles to the no particle state of inertial observer (called Minkowski vacuum). The decoherence effect, produced by the Unruh effect, suppresses the QRs such as QC \cite{J. Wang}, quantum discord \cite{E. G. Brown,J. Doukas} and entanglement \cite{J. Doukas} in the case of bosonic or Dirac field modes. The degradation of QRs unavoidably decreases the confidence of some quantum information tasks like quantum teleportation. In this context it is really important to preserve QRs from decoherence during the teleportation process. Here we investigate the teleportation of single and two-qubit quantum states in which the resource state of the teleportation is affected by  the Unruh effect experienced by a mode of a free Dirac field, as seen by a relativistically accelerated observer. This kind of the teleportation channel discussed in this paper is called Unruh noise channel throughout the text.

In addition to the teleportation of the whole quantum state, we also investigate the teleportation of the information encoded into a particular parameter. In contrast to quantum state teleportation where the quality of teleportation is characterized by fidelity, the credibility of teleportation of specific information is usually determined by QFI \cite{C. W. Helstrom,Holevo,M. G. Genoni,Lu}. Representing the sensitivity of the state with respect to changes in a parameter, it plays an important role in parameter estimation theory and is extensively employed in QIP. In particular, the QFI has many applications in quantum information tasks such as entanglement detection \cite{N. Li,Farajollahi B.}, specifying the non-Markovianity \cite{X. M. Lu,Rangani Jahromi H.,Rangani Jahromi H2.}, quantum thermometry~\cite{HRJ20}, and consideration of uncertainty relations \cite{P. Gibilisco,P. Gibilisco2,Y. Watanabe}. Hence it is of interest to study the QFI in relativistic framework. Nevertheless, it is shown that the QFI is fragile and can be broken easily because of unavoidable decoherence effects \cite{J. Kolodynski,J. Ma,K. Berrada,Y. M,Y. L}. This is the most restricting factor in QFI applications for quantum teleportation. Therefore, protecting the QFI from decoherence is a fundamental subject.

In PM, associated with a positive-operator valued measure (POVM), the system state does not completely collapse  such that the initial state could be reversed with some operations. Recently, PM together with PMR have been exploited as a practical method to protect quantum correlations of two-qubits as well as two-qutrits and the fidelity of a single-qubit, from amplitude damping (AD) decoherence \cite{Q. Q. Sun,Z. X. Man,X. Xiao3,X. Xiao4,Rangani Jahromi H3}. In ref. \cite{X. Xiao2} the effect of PMs (hereafter we use PMs to indicate both PM and PMR) on QFI of a teleported single-qubit state under the AD noise has been studied and illustrated that the combination of PM and PMR could totally eliminate the influence of decoherence. The effects of PM and PMR on the enhancement of quantum coherence and QFI, transmitted under a quantum spin-chain channel, have been considered in ref. \cite{Z. Liu}. Moreover it has been shown that PM and PMR are able to improve the fidelity of teleportation when one or both qubits of the maximally entangled state shared between Alice and Bob suffer from the AD decoherence \cite{T. Pramanik}. It was also shown in ref \cite{T. Pramanik} that this protocol works for the Werner states.
However limited attention has been paid to protect the QRs and QFI against Unruh decoherence during the procedure of teleportation. Motivated by this, we study the enhancement effect of PM and PMR on QRs and QFI of the teleported state through the Unruh noise channel for both single- and two-qubit input quantum states.

In this paper, we have investigated the following scenario: the system consists of an inertial observer Alice and a uniformly accelerated observer Rob. Two PMs are performed before and after Rob’s acceleration, which are called PM and PMR, respectively. Then we use the above mentioned system as a resource in order to teleport single- and two-qubit states, and consider how the degradation effect of the Unruh channel on QRs and QFI of the teleported sate as well as teleportation fidelity can be improved by PM or PMR. According to our results, the combined effect of PM and PMR with the same strengths ($ p=q $) may improve QRs and QFI with respect to phase parameter $ \varphi $, of the teleported state, and also teleportation fidelity in both single-qubit and two-qubit scenarios.
Our study differs from~\cite{YXG+14} in which the performance of the QFI of an arbitrary two-qubit state under Unruh effect has been addressed. Here the two-qubit state exposed to the Unruh noise in the Dirac field and affected by PMs, is used as a \emph{resource} to teleport single- and two-qubit unknown quantum states.
Our work also differs from \cite{Jafarzadeh M.} in which the two-qubit state exposed to the Unruh noise in the scalar field has been used as a resource to teleport a two-qubit state. 

This paper is organized as follows: In Sec.~\ref{Preliminaries } we give a brief description about teleportation, PM, PMR, QRs and QFI. The physical model is presented in Sec.~\ref{Model}. The probability of preparing the resource state of the quantum channel in teleportation protocol is discussed in Sec.~\ref{prob}. We study the single-qubit teleportation as well as two-qubit teleportation under the Unruh noise channel in Sec.~\ref{1qbtele} and Sec.~\ref{2qbtele}, respectively. Finally, Sec.~\ref{conclusion} is devoted to conclusion.

\section{PRELIMINARIES  \label{Preliminaries }}  
\subsection{Teleportation}
The main idea of quantum teleportation is transferring quantum information about an unknown quantum state in one location (Alice) to another location (Bob) where it is spatially separated. An important factor in quantum teleportation is the channel connecting sender and receiver. In standard teleportation protocol $ T_{0} $, local quantum operations, used to teleport the quantum state, includes Bell measurements and Pauli rotations. According to Bowen and Bose results, the standard teleportation protocol $ T_{0} $ with mixed states as resource is tantamount to a generalized depolarizing channel \cite{G. Bowen}. 
\subsubsection{Single-qubit teleportation}
As mentioned above, teleportation protocol using a two-qubit mixed state as a resource, acts as a generalized depolarizing channel $\Lambda\left( \rho_{\text{ch}}\right)$.
Therefore, for an arbitrary single-qubit state to be teleported (henceforth called input state $\rho_{\text{in}}$),
the output state $\rho_{\text{out}}$ of the teleportation is obtained as follows \cite{G. Bowen}
\begin{eqnarray}\label{a1}
\nonumber\rho_{\text{out}}=\Lambda\left( \rho_{\text{ch}}\right)\rho_{\text{in}},~~~~~~~~~~~~~\\
=\sum^{3}_{i=0}\text{Tr}\left(\mathcal{B}_{i}\rho_{\text{ch}} \right)\sigma_{i}\rho_{\text{in}}\sigma_{i}, 
\end{eqnarray}
in which $ \mathcal{B}_{i} $'s represent the Bell states associated with the Pauli matrices $ \sigma_{i} $'s by
\begin{equation}\label{a2}
\mathcal{B}_{i}=\left(\sigma_{0}\otimes\sigma_{i}\right)\mathcal{B}_{0}\left(\sigma_{0}\otimes\sigma_{i}\right); \;i=1,2,3 
\end{equation}
where $ \sigma_{0}=I $, $ \sigma_{1}=\sigma_{x} $, $ \sigma_{2}=\sigma_{y} $, $ \sigma_{3}=\sigma_{z} $, and $I$ is the $2\times 2$ identity matrix.
We have $ \mathcal{B}_{0}=\frac{1}{2}\left(|00\rangle +|11\rangle\right) \left(\langle00| +\langle11| \right)  $, without loss of the generality.
Besides, 
	$\rho_\text{ch}$ indicates the resource state of the quantum channel shared between Alice and Bob. It should be noted that this resource state of the teleportation channel  will be  affected by the Unruh effect in our relativistic model. 
\subsubsection{Two-qubit teleportation}
\begin{figure}[ht]
	\centering
	\includegraphics[width=8cm]{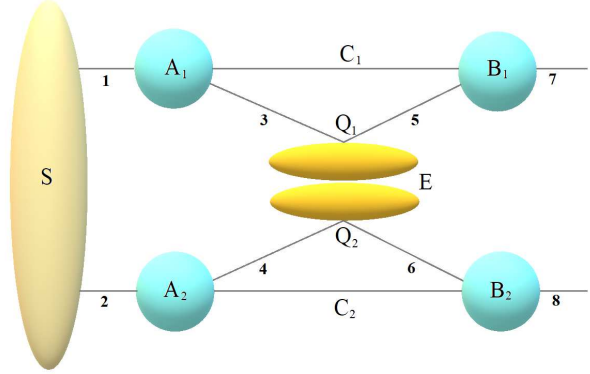}
	\caption{ Schematic drawing of entanglement teleportation.
		An unknown entangled state $ \rho_{\text{in}} $ is generated by source $S$, and its particles are dispensed into $A_{1}$ and $A_{2}$. Besides, two independent entangled pairs, numbered (3,5) and (4,6), are produced from source $E$. These pairs, each characterized by density matrix $ \rho_{\text{ch}} $, play the role of the quantum channels $Q_1$ and $Q_2$. The measurement result at $A_{i}$ ($i=1,2$) is transmitted through the classical channel $C_{i}$ to $B_{i}$. Based on the information received by the classical communication, the unitary transformations are performed on the particles 5 and 6 at $B_{i}$ ($i=1,2$) to complete the teleportation.
	}
	\label{teleportation}
\end{figure}

Teleportation of an unknown entangled state via two independent quantum channels has been studied by Lee and Kim \cite{Lee J}. Actually, their protocol may be carried out by doubling the standard teleportation protocol $ T_{0} $. Figure \ref{teleportation} displays the schematic drawing of entanglement teleportation.  

Generalizing equation (\ref{a1}), the output state of the entanglement teleportation is found as follows
\begin{equation}\label{a3}
\rho_{\text{out}}=\sum_{ij}p_{ij}\left( \sigma_{i}\otimes\sigma_{j}\right) \rho_{\text{in}}\left( \sigma_{i}\otimes\sigma_{j}\right),\  i,j=0,x,y,z.
\end{equation}
where $ p_{ij}=\text{Tr}\left(\mathcal{B}_{i}\rho_{\text{ch}} \right)\text{Tr}\left(\mathcal{B}_{j}\rho_{\text{ch}} \right) $ and $ \sum p_{ij}=1 $. 

\subsection{Partial measurement (PM) and partial measurement reversal(PMR) \label{PM}}

We first give a brief introduction of the PM and PMR. In contrast to the standard von Neumann projective measurement, which completely collapses the measured system, PM, as a generalization of standard von Neumann projective measurement, does not totally collapse the initial state into the eigenstates, and hence the measured state is reversible by proper PMR operations. For a single-qubit, the PM is described by the following pair of measurement operators:
\begin{equation}\label{a5}
M_{0}=\sqrt{p}|0\rangle\langle0|,
\end{equation}
\begin{equation}\label{a4}
M_{1}=\sqrt{1-p}|0\rangle\langle0|+|1\rangle\langle1|,
\end{equation}
where $ p\left( 0\leq p\leq1\right)  $ is the strength of PM and $ M_{0}^{\dagger}M_{0}+M_{1}^{\dagger}M_{1}=I $. $ M_{0} $ is identical to von Neumann projective measurement and is irreversible, while $ M_{1} $ is a PM which could be reversed for the case $p\neq1$. On the other hand, the PMR can be described by a non-unitary operator
\begin{equation}\label{a6}
M_{1}^{-1}=\frac{1}{\sqrt{1-q}}\left(\begin{array}{cc}
0& 1 \\
1 &0 \\
\end{array}\right)\left(\begin{array}{cc}
\sqrt{1-q}& 0 \\
0 & 1 \\
\end{array}\right)\left(\begin{array}{cc}
0& 1 \\
1 &0 \\
\end{array}\right)=\frac{1}{\sqrt{1-q}}XM_{1}X,
\end{equation}
where $ X=|0\rangle\langle1|+|1\rangle\langle0| $ is the bit-flip operation. Last term of Eq.(\ref{a6}) implies that the PMR can be implemented  physically by the sequence of a bit-flip operation, another PM with measurement strength $ q $, and a second bit-flip operation.
Therefore, the PMR could exactly undo the PM by choosing $q=p$. 

\subsection{Quantum Fisher information\label{QFI}}
QFI is an important concept in parameter estimation theory. QFI of an unknown parameter $ \lambda $ encoded in quantum state $ \rho\left(\lambda \right) $ is defined as \cite{C. W. Helstrom,Braunstein SL}
\begin{equation}\label{01}
F_{Q}\left( \lambda\right)=\text{Tr}\left[\rho\left(\lambda \right)L^{2} \right]=\text{Tr}\left[\left( \partial_{\lambda}\rho\left(\lambda \right)\right) L\right], 
\end{equation}
where L, the symmetric logarithmic derivative (SLD), is given by $ \partial_{\lambda}\rho\left(\lambda \right)=\frac{1}{2}\left(L\rho\left(\lambda \right)+\rho\left(\lambda \right)L\right), $ with $ \partial_{\lambda}=\partial/\partial\lambda $.

A simple and explicit expression can be acquired for the single-qubit state. Any qubit state can be expressed in the Bloch sphere representation as $\rho=\frac{1}{2}\left(I+\boldsymbol{\omega}\cdot\boldsymbol{\sigma}\right)$,
where $ \boldsymbol{\omega}=\left(\omega_{x},\omega_{y},\omega_{z} \right)^{T} $ is the Bloch vector and $ \sigma=\left(\sigma_{x},\sigma_{y},\sigma_{z} \right)  $ indicates the Pauli matrices. Hence the QFI of the single-qubit state can be formulated as follows \cite{W. Zhong}

\begin{equation}\label{a12}
F_{Q}\left( \lambda\right)=\left\{\begin{array}{cc}
|\partial_{\lambda}\boldsymbol{\omega}|^{2}+\frac{\left( \boldsymbol{\omega}\cdot\partial_{\lambda}\boldsymbol{\omega}\right) ^{2}}{1-|\boldsymbol{\omega}|^{2}},&~~~~~~~~~~ |\boldsymbol{\omega}|<1, \\
|\partial_{\lambda}\boldsymbol{\omega}|^{2},~~~~~~~~~~~~~~~~~~~ & ~~~~~~~~~~|\boldsymbol{\omega}|=1. \\
\end{array}\right.
\end{equation}
where $ |\boldsymbol{\omega}|<1 $ is used for a mixed state while $ |\boldsymbol{\omega}|=1 $ is applicable for a pure state. 

\subsection{Quantum resources \label{QR}}
\emph{Quantum coherence}.
QC arising from the superposition principle is an important resource in quantum information and quantum computation processing. It plays a fundamental role in quantum mechanics. Various measures are expressed to quantify the coherence such as, trace norm distance coherence \cite{L. H. Shao}, $ l_{1} $ norm, and relative entropy of coherence \cite{Baumgratz T}. For a quantum state with the density matrix $ \rho $, the $ l_{1} $ norm measure of quantum coherence \cite{Baumgratz T} quantifying the coherence through the off diagonal elements of the density matrix in the reference basis, is given by

\begin{equation}\label{a7}
\mathcal{C}_{l_{1}}\left(\rho \right)=\sum_{i,j\atop i\ne j}|\rho_{ij}|,
\end{equation}

\emph{Entanglement}. Entanglement is recognized as a resource in quantum information processing (QIP) and is accountable to the advantage of many quantum computation and communication tasks. Actually, entanglement indicates correlations regarding non separability of the state of a composite quantum system. Entanglement of a bipartite system is quantified conveniently by concurrence \cite{Wootters WK} which can be computed analytically for a X state as follows

\begin{equation}\label{a8}
C(\rho)=2\text{max}\left\lbrace 0,C_{1}(\rho),C_{2}(\rho) \right\rbrace ,
\end{equation}
where $ C_{1}(\rho)=|\rho_{14}|-\sqrt{\rho_{22}\rho_{33}}, C_{2}(\rho)=|\rho_{23}|-\sqrt{\rho_{11}\rho_{44}} $, and $ \rho_{ij} $'s are the elements of density matrix. Concurrence equals unity for maximally entangled states and vanishes for separable states.

\emph{Quantum discord}.
Quantum discord representing quantumness of the state of quantum system is a resource for certain quantum technologies. It can be preserved for a long time even when entanglement shows a sudden death. QD for any bipartite system is defined as difference between total correlations (i.e., quantum mutual information) and classical correlations. Computation of QD for general states is not usually a convenient task since it involves the optimization of the classical correlations. However, for a two-qubit $ X $ state system, the analytical expression of QD can be obtained as \cite{Wang C-Z}

\begin{equation}\label{a9}
QD(\rho_{AB})=\text{min}\left(Q_{1},Q_{2} \right),
\end{equation}
where
\begin{eqnarray}\label{a10}
\nonumber Q_{j}=H\left(\rho_{11} +\rho_{33}  \right)+\sum_{i=1}^{4}\lambda_{i}\text{log}_{2}\lambda_{i}+D_{j},~~~\left(j=1,2\right),~~~~~\\
D_{1}=H\left(\frac{1+\sqrt{\left[1-2\left( \rho_{33} +\rho_{44}\right)  \right] ^{2}+4\left(|\rho_{14}|+|\rho_{23}| \right) ^{2}}}{2} \right), \\\nonumber
D_{2}=-\sum_{i}\rho_{ii}\text{log}_{2}\rho_{ii}-H\left(\rho_{11} +\rho_{33}  \right),~~~~~~~~~~~~~~~~~~~~~~~~~~\\\nonumber
\nonumber H\left(x \right)=-x\text{log}_{2}x-\left(1-x \right)\text{log}_{2}\left(1-x \right),~~~~~~~~~~~~~~~~~~~~~~~~  
\end{eqnarray}
and $ \lambda_{i} $'s denote the eigenvalues of density matrix $ \rho_{AB} $.

\begin{figure}[ht]
	\centering
	\includegraphics[width=7cm]{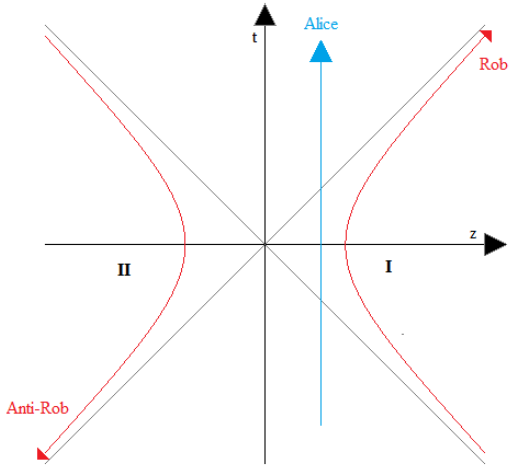}
	\caption{\small Rindler spacetime embedded into Minkowski spacetime:
			a uniformly accelerated observer (Rob) with acceleration $a$
			travels on a hyperbola constrained to region I, while a fictitious
			observer (anti-Rob) travels on a corresponding hyperbola in region II. The straight line indicates the world line of an inertial observer (Alice).}
	\label{f1}
\end{figure}

\section{Physical model \label{Model}}

We consider a system including an inertial observer Alice (A) and a uniformly accelerated observer Rob (R).
Let Alice and Rob initially share the following entangled state of two Dirac field modes
\begin{equation}\label{initial state}
|\Psi\left( 0\right) \rangle=\text{sin}~\frac{\vartheta}{2}|0\rangle_{A}|0\rangle_{R}+\text{cos}~\frac{\vartheta}{2}|1\rangle_{A}|1\rangle_{R},
\end{equation}
where $\ket{0}$ and $\ket{1}$ denote the Minkowski vacuum and excited states~\cite{YXG+14}, respectively.

We assume that Rob first performs a PM of the form (\ref{a4}) on his particle. State (\ref{initial state}) reduces to

\begin{equation}\label{Psi1}
|\Psi\left( 1\right) \rangle=\frac{1}{\sqrt{N_{1}}}\left( \text{sin}~\frac{\vartheta}{2}~\sqrt{\overline{p}}|0\rangle_\text{A}|0\rangle_\text{R}+\text{cos}~\frac{\vartheta}{2}|1\rangle_\text{A}|1\rangle_\text{R}\right) ,
\end{equation}
where $ N_{1}=\text{sin}^{2}\frac{\vartheta}{2}~\overline{p}+\text{cos}^{2}\frac{\vartheta}{2} $ represents the normalization factor and $ \overline{p}=1-p $.
If $ p $ approaches unity, state (\ref{Psi1}) is projected into  state $ |1_{A}\rangle|1_{R}\rangle $, which is not affected by the Unruh effect.

Now Rob begins to accelerate uniformly. Hence he sees that  the Unruh effect is experienced by the Dirac mode which is in the
vacuum state from an inertial perspective such that
his detector perceives a Fermi–Dirac
distribution of particles. Rindler coordinates $ \left(\tau,\xi\right)  $ is a better option to describe what he sees.  

 Rindler spacetime manifests two regions I and II, causally disconnected (Fig.\ref{f1}). Because of the eternal acceleration, Rob travels on a hyperbola compelled in the region I. 

In Rob's reference frame the Minkowski vacuum state can be expressed in terms of the Rindler regions I and  II states \cite{P. M. Alsing}:

\begin{equation}\label{2}
\ket{0}_{M}=\text{cos}~r|0\rangle_{\text{I}}|0\rangle_{\text{II}}+\text{sin}~r|1\rangle_{\text{I}}|1\rangle_{\text{II}},
\end{equation}
while the excited state appears as a product state
\begin{equation}\label{3}
\ket{1}_{M}=|1\rangle_{\text{I}}|0\rangle_{\text{II}},
\end{equation}
where $ r=\text{arccos}\sqrt{1+e^{\frac{-2\pi\omega}{a}}} $, $ \omega $ is the Dirac particle frequency and $ a $ is the acceleration. Since $ 0< a<\infty $, therefore $ r\in\left[ 0,\pi/4\right]$.
Note that the observers in regions I and II are causally disconnected. Since the mode corresponding to II in not observable for Rob in region I, it should be traced out. 

Expanding  Minkowski particle states $\ket{0}_{R}$ and $\ket{1}_{R}$ by using, respectively, Eqs.~(\ref{2}) and (\ref{3}), we find that the state represented in (\ref{Psi1}) changes to
\begin{equation}\label{a14}
|\Psi\left( 2\right) \rangle=\frac{1}{\sqrt{N_{1}}}\left[ \text{sin}~\frac{\vartheta}{2}~\sqrt{\overline{p}}\left(\text{cos}~r|0\rangle_\text{A}|0\rangle_{\text{I}}|0\rangle_{\text{II}}+\text{sin}~r|0\rangle_\text{A}|1\rangle_{\text{I}}|1\rangle_{\text{II}} \right) +\text{cos}~\frac{\vartheta}{2}|1\rangle_\text{A}|1\rangle_{\text{I}}|0\rangle_{\text{II}} \right]. 
\end{equation}
In the next step, a PMR is carried out by Rob in the region I. If the PMR is successfully performed, we achieve
\begin{align}
\label{a15}
\nonumber \ket{\Psi\left(3\right)}=&\frac{1}{\sqrt{N_{2}}}\Bigg[\Bigg. \text{sin}~\frac{\vartheta}{2}~\sqrt{\overline{p}}\left(\text{cos}~r\ket{0}_\text{A}\ket{0}_{\text{I}}\ket{0}_{\text{II}}+\text{sin}~r\sqrt{\overline{q}}\ket{0}_\text{A}\ket{1}_{\text{I}}\ket{1}_{\text{II}} \right)\\
&+\text{cos}~\frac{\vartheta}{2}\sqrt{\overline{q}}\ket{1}_\text{A}\ket{1}_{\text{I}}\ket{0}_{\text{II}}\Bigg.\Bigg],
\end{align}
where $N_{2}=\text{sin}^{2}\frac{\vartheta}{2}~\overline{p}\;\text{cos}^{2}r+\text{sin}^{2}\frac{\vartheta}{2}~\overline{pq}~\text{sin}^{2}r+\text{cos}^{2}\frac{\vartheta}{2}~\overline{q} $ is the normalization factor and $ \overline{q}=1-q $, in which $ q $ represents the second PM strength. Since Rob is restricted to region I due to the causality condition, we trace the state over region II, resulting in the following mixed state between Alice and Rob \cite{X. Xiao}
\begin{eqnarray}\label{a16}
\rho_{A,R}=\frac{1}{N_{2}}\left(\begin{array}{cccc}
\text{sin}^{2}\frac{\vartheta}{2}~\overline{p}~\text{cos}^{2}r& 0& 0 & \text{sin}~\frac{\vartheta}{2}\text{cos}~\frac{\vartheta}{2}~\sqrt{\overline{pq}}~\text{cos}r\\
0 & \text{sin}^{2}\frac{\vartheta}{2}~\overline{pq}~\text{sin}^{2}r&0 & 0 \\
0 & 0& 0 &0 \\
\text{sin}~\frac{\vartheta}{2}\text{cos}~\frac{\vartheta}{2}~\sqrt{\overline{pq}}~\text{cos}r & 0 & 0& \text{cos}^{2}\frac{\vartheta}{2}~\overline{q} \\
\end{array}\right).\\\nonumber
\end{eqnarray}
Next we discuss the preparation probability of  state (\ref{a16}).

\section{Preparation probability}\label{prob}
In this section, we address the probability of preparing the system in state (\ref{a16}).
If the state of a quantum system is $\rho$ immediately	before the measurement then the probability that result $m$ occurs is given by $P_{m}=\text{tr}\left(M^\dagger_m M_m\rho\right)$ \cite{MI00}, where $m$ is the measurement outcome.
In order to find the preparation probability, we start with the initial state (\ref{initial state}) shared between Alice and Rob.
If the state of the system is $\ket{\Psi\left(0\right)}$ immediately before the first PM then the probability that $M_0$ is carried out successfully is given by
\begin{align}
\nonumber P_1=&\bra{\Psi\left(0\right)}M_{1}^{\dagger}M_{1}\ket{\Psi\left( 0\right)}\\
=&\text{sin}^{2}\frac{\vartheta}{2}~\overline{p}+\text{cos}^{2}\frac{\vartheta}{2}.
\end{align}
Now we proceed with finding the probability that the PMR is carried out successfully. As we explained in section \ref{PM}, the PMR can be implemented by concatenation of a bit-flip operation, second PM of strength $q$, and a final bit-flip operation. Assuming that the first PM has been performed successfully, we find that the state of the system is
$\rho=\text{tr}_{\text{II}}\left({X\ket{\Psi\left(2\right)}\bra{\Psi\left(2\right)}X}\right)$ immediately before the second PM, and hence the success probability of the second PM is obtained as follows
\begin{align}
\nonumber	P_2=&\text{tr}\left(M_{1}^{\dagger}M_{1}\rho\right)\\
=&\frac{\text{sin}^{2}\frac{\vartheta}{2}\;\overline{p}\;\text{cos}^{2}r+\text{sin}^{2}\frac{\vartheta}{2}~\overline{pq}~\text{sin}^{2}r+\text{cos}^{2}\frac{\vartheta}{2}~\overline{q}}{\text{sin}^{2}\frac{\vartheta}{2}~\overline{p}+\text{cos}^{2}\frac{\vartheta}{2}}.
\end{align}
Now we are ready to find the preparation probability of the system in state (\ref{a16}), i.e., the probability that the first and the second PMs are carried out successfully:    
\begin{align}
\nonumber P=&P_{1}.P_{2}\\
=&\text{sin}^{2}\frac{\vartheta}{2}\;\overline{p}\;\text{cos}^{2}r+\overline{q}\left(\text{sin}^{2}\frac{\vartheta}{2}\overline{p}\;\text{sin}^{2}r+\text{cos}^{2}\frac{\vartheta}{2} \right). 
\end{align}

\begin{figure}[ht]
	\centering
	\includegraphics[width=8cm]{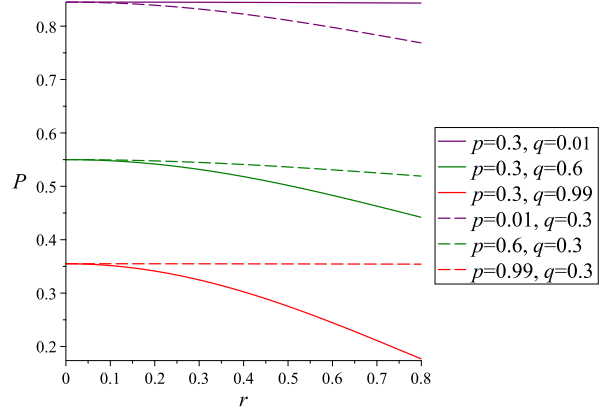}
	\caption{\small The probability of preparing the state of the channel as a function of $r$ for different values of $p$ and $q$, fixing $\vartheta=\pi/2 $.}
	\label{F0}
\end{figure}

In Fig. \ref{F0}, the preparation probability is plotted versus acceleration parameter $ r $, for different values of $p$ and $q$, fixing $ \vartheta=\pi/2 $. 
We see that the preparation probability decreases with increase in $p$ or $q$. Moreover, it is more robust against the increase of $r$ for $ p>q $ than the case $ q>p $. 
The same results are obtained for $ 0<\vartheta<\frac{\pi}{2} $.

\noindent Next we discuss how PM and PMR affect the  QRs and QFI of the  state teleported  through the Unruh noise channel. 

\section{Single-qubit teleportation under the Unruh noise channel\label{1qbtele}}
\par In this section, the QFIs and QC of the teleported single-qubit state, through the Unruh noise channel are investigated. We consider $|\psi_{in}\rangle=\text{cos}~\theta/2 |0\rangle+e^{i\varphi}~\text{sin}~\theta/2 |1\rangle, 0\leq\theta\leq\pi,\ 0\leq\varphi\leq2\pi$ as the input state in the process of teleportation, where $ \theta $ and $ \varphi $ are the weight and phase parameters, respectively. We use the shared state between Alice and Rob, Eq. (\ref{a16}), as the resource ($ \rho_{A,I}=\rho_{ch} $) to teleport the single-qubit input state. Using Eq. (\ref{a1}), the output state can be obtained as follows
\begin{eqnarray}\label{a17}
\rho_{\text{out}}^{\text{PM}}=\frac{1}{N_{2}}\left(\begin{array}{cc}
\mathcal{A}\text{cos}^{2}\frac{\theta}{2}+\mathcal{D}\text{sin}^{2}\frac{\theta}{2} & \mathcal{F}e^{-i\varphi}\text{sin}~\theta\\
\mathcal{F}e^{i\varphi}\text{sin}~\theta & \mathcal{A}\text{sin}^{2}\frac{\theta}{2}+\mathcal{D}\text{cos}^{2}\frac{\theta}{2}\\
\end{array}\right),\\\nonumber
\end{eqnarray}
where
\begin{eqnarray}\label{a18}
\nonumber \mathcal{A}=\text{sin}^{2}\frac{\vartheta}{2}~\overline{p}~\text{cos}^{2}r+\text{cos}^{2}\frac{\vartheta}{2}~\overline{q},\\
\mathcal{D}=\text{sin}^{2}\frac{\vartheta}{2}~\overline{pq}~\text{sin}^{2}r,~~~~~~~~~~~~~\\\nonumber
\mathcal{F}=\text{sin}\frac{\vartheta}{2}\text{cos}\frac{\vartheta}{2}\sqrt{\overline{pq}}~\text{cos}~r,~~~~~~~\\\nonumber
\end{eqnarray}
\par The QFIs of  input state $ |\psi_{in}\rangle=\text{cos}~\theta/2 |0\rangle+e^{i\varphi}~\text{sin}~\theta/2 |1\rangle$,   with respect to parameters $ \theta $ and $ \varphi $ are easily found to be $ F_{\text{in}}\left(\theta \right)=1 $ and $ F_{\text{in}}\left(\varphi \right)=\text{sin}^{2}\theta $, respectively. It is seen that $ F_{\text{in}}\left(\varphi \right) $ is dependent on $ \theta $ and is maximized for $ \theta=\frac{\pi}{2} $ while $ F_{\text{in}}\left(\theta \right) $ is independent of weight parameter $ \theta $ and has a constant value. Therefore, if the input state is not exposed to the Unruh noise channel, it is preferred to be balance-weighted to achieve the best estimation  of the phase parameter.

\par We use Eq.~(\ref{a12}) to calculate the QFIs of the output state (the teleported state) (\ref{a17}). The corresponding Bloch vector is given by
\begin{equation}
\boldsymbol{\omega}=\frac{1}{N_2}\left(2\mathcal{F}\text{sin}\theta~\text{cos}\varphi,2\mathcal{F}\text{sin}\theta~\text{sin}\varphi,(\mathcal{A}-\mathcal{D})~\text{cos}\theta\right),
\end{equation}
where its components are calculated by $\omega_i=\text{Tr}(\rho_{\text{out}}^{\text{PM}}\sigma_i),\;i=1,2,3$.  
Therefore, the QFIs with respect to weight and phase parameters are found, respectively, as follows
\begin{equation}
F_{\text{out}}^{\text{PM}}\left(\theta \right)=\frac{1}{N_{2}^{2}}\left(\left(\mathcal{A}-\mathcal{D} \right) ^{2}\text{sin}^{2}\theta+4\mathcal{F}^{2}\text{cos}^{2}\theta+\frac{\frac{1}{4}\left( \left(\mathcal{A}-\mathcal{D} \right) ^{2}-4\mathcal{F}^{2}\right) ^{2}\text{sin}^{2}2\theta}{N_{2}^{2}-\left(\mathcal{A}-\mathcal{D} \right) ^{2}\text{cos}^{2}\theta-4\mathcal{F}^{2}\text{sin}^{2}\theta} \right),      
\end{equation}
\begin{equation}\label{EQ54}
F_{\text{out}}^{\text{PM}}\left(\varphi \right)=4|\frac{\mathcal{F}\text{sin}~\theta}{N_{2}}|^{2}.     
\end{equation}
\begin{figure}[ht!]
	\centering
	\subfigure[]{\includegraphics[width=6.6cm]{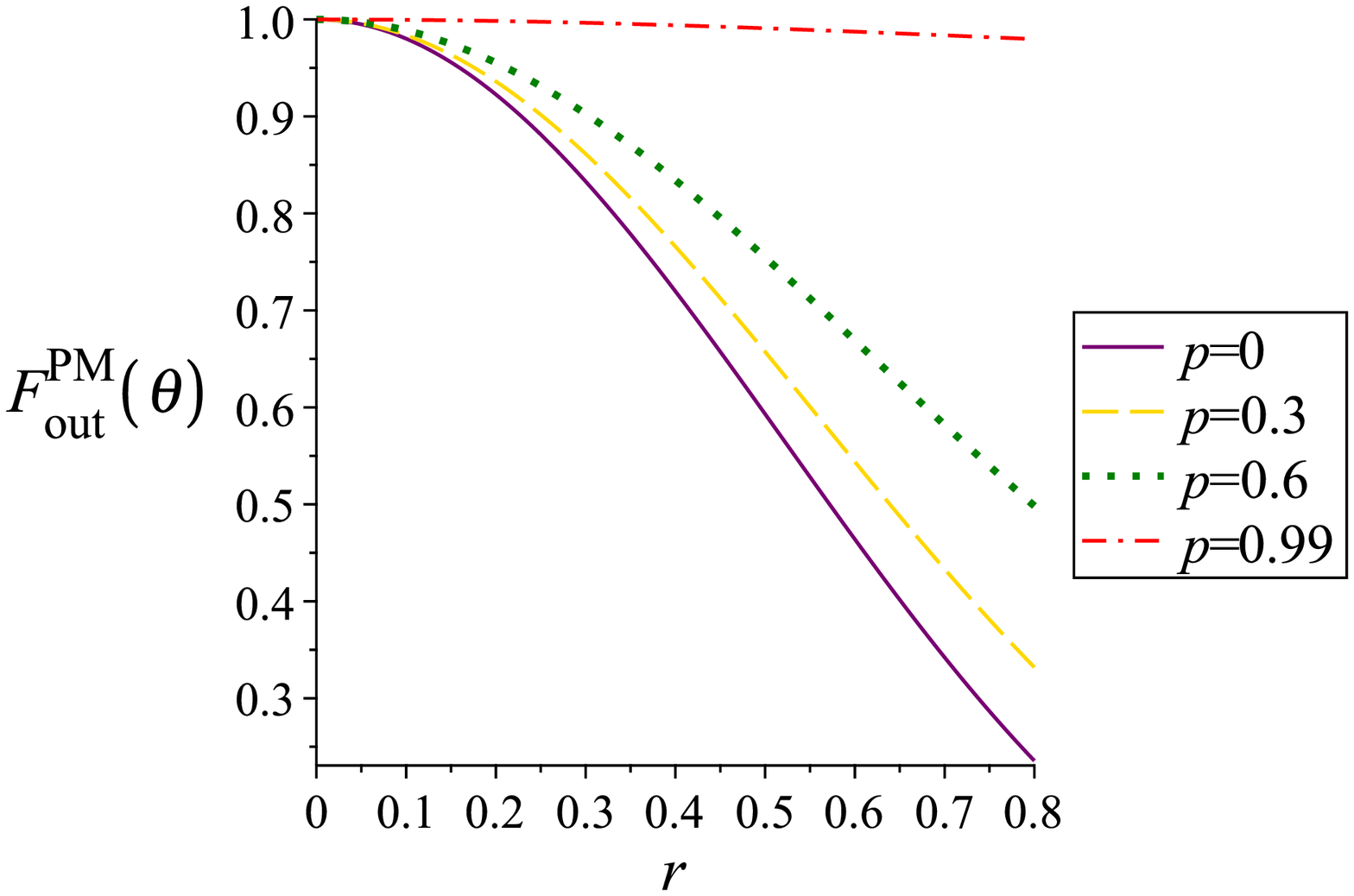}\label{ff4} }
	\subfigure[]{\includegraphics[width=6.6cm]{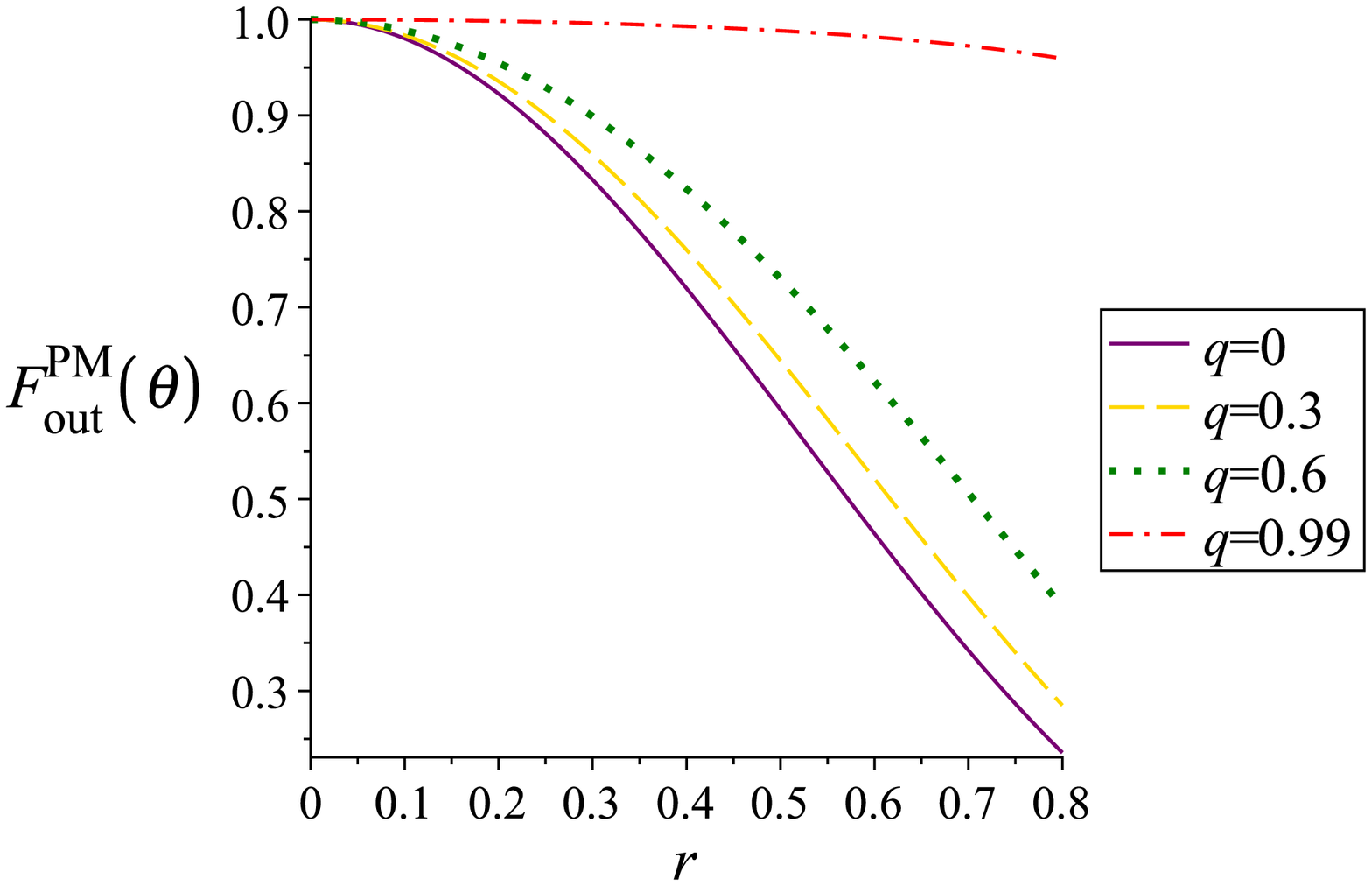}\label{ff5} }
	
	\caption{\small QFI of the single-qubit teleported state with respect to the weight parameter, $ \theta $, as a function of acceleration parameter $ r $ by fixing $ \theta=\frac{\pi}{2} $ and for $ 0<\vartheta<\pi $ for (a) $ q=0 $, (b) $ p=0 $.}
	\label{FF}   
\end{figure} 
\par In Fig. \ref{FF}, the QFI with respect to the weight parameter, $ F_{\text{out}}^{\text{PM}}\left(\theta \right) $, for single-qubit state teleportation through the pure Unruh decoherence and for the case that the combination of PM and PMR have been applied, is plotted as a function of acceleration parameter $ r $. It can be seen that after teleportation under pure Unruh channel (i.e., $ p=q=0 $), when the acceleration increases, the QFI decays monotonously for all values of the initial parameter $ \vartheta $. Studying the behavior of $ F_{\text{out}}^{\text{PM}}\left(\theta \right) $, when the PM and PMR are applied on the channel, we observe that applying either PM (i.e., $ q=0 $) or PMR (i.e., $ p=0 $) may improve $ F_{\text{out}}^{\text{PM}}\left(\theta \right) $ for all initial states of the channel (see Figs. \ref{ff4} and \ref{ff5} ). For sufficiently strong measurement strength ($ p\rightarrow1 $ or $ q\rightarrow1 $), the precision of estimating weight parameter can be enhanced remarkably and it is almost robust against the Unruh decoherence. 

\begin{figure}[ht]
	\centering
	\includegraphics[width=6.7cm]{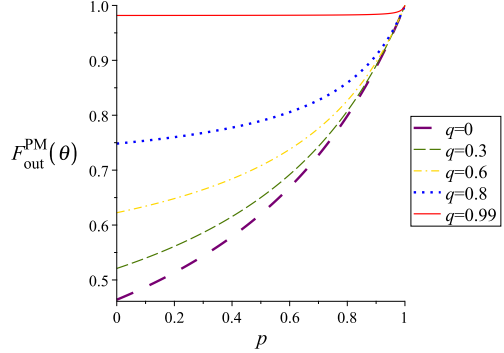}
	\caption{\small QFI of the single-qubit teleported state with respect to the weight parameter, $ \theta $, as a function of PM strength, $p$, for $r = 0.6$, and different values of PMR strength.}
	\label{qf}
\end{figure}

\par The important question that comes up is that if the acceleration is constant, how one can control the QFI by applying PM and PMR. In Fig. \ref{qf} we consider the $ F_{\text{out}}^{\text{PM}}\left(\theta \right) $  behavior versus $ p $. It is observed that in the absence of PMR (i.e., $q=0$), $ F_{\text{out}}^{\text{PM}}\left(\theta \right) $ enhances with increase in $ p $ (space dashed purple line) for all values of the channel parameter $ \vartheta $. It is also seen that with the combined effect of PM and PMR, estimation precision of weight parameter is also improved. We obtain the same results investigating the behavior of $ F_{\text{out}}^{\text{PM}}\left(\theta \right) $ versus $ q $.

\begin{figure}[ht!]
	\centering
	\subfigure[]{\includegraphics[width=6.6cm]{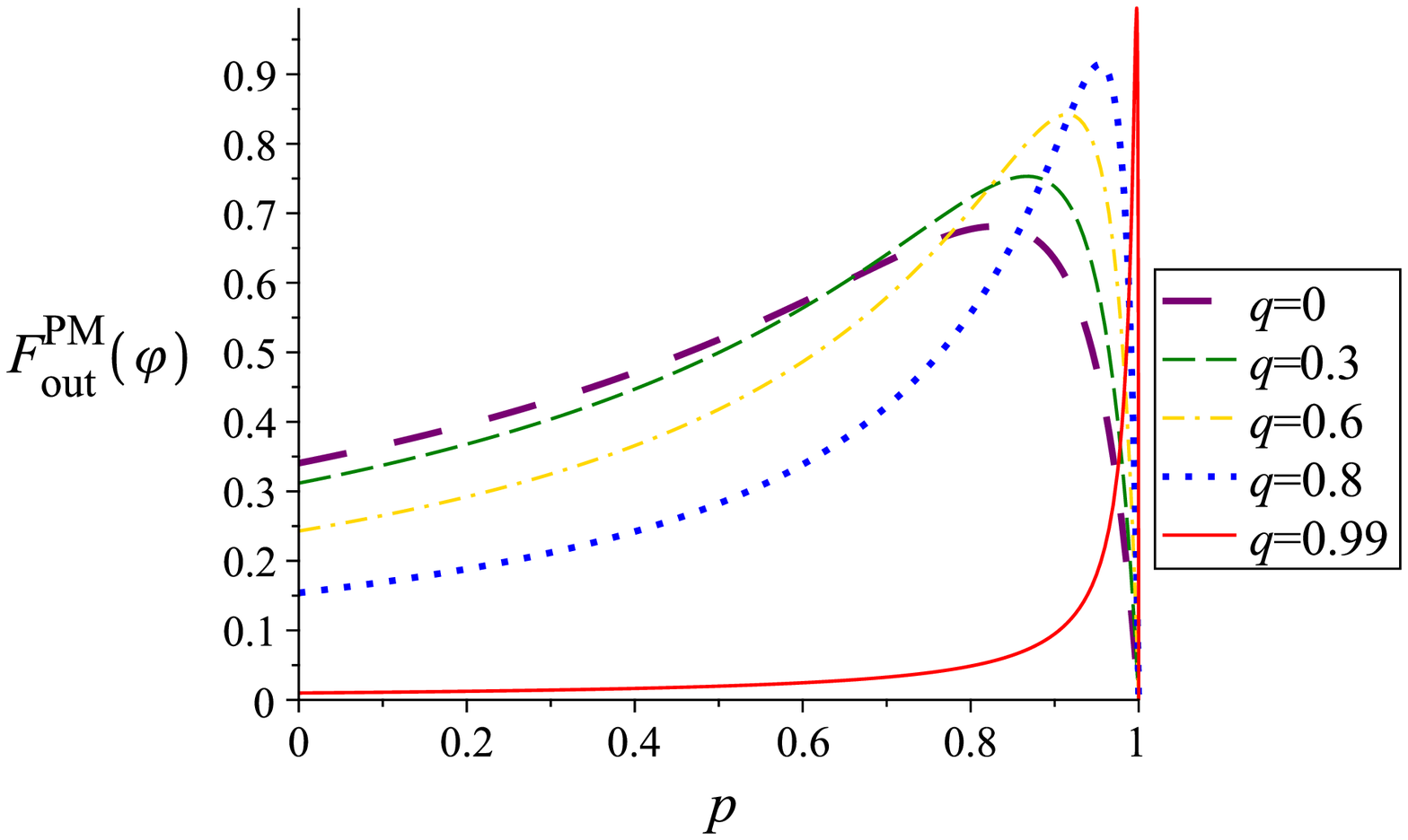}\label{qf21} }
	\subfigure[]{\includegraphics[width=6.6cm]{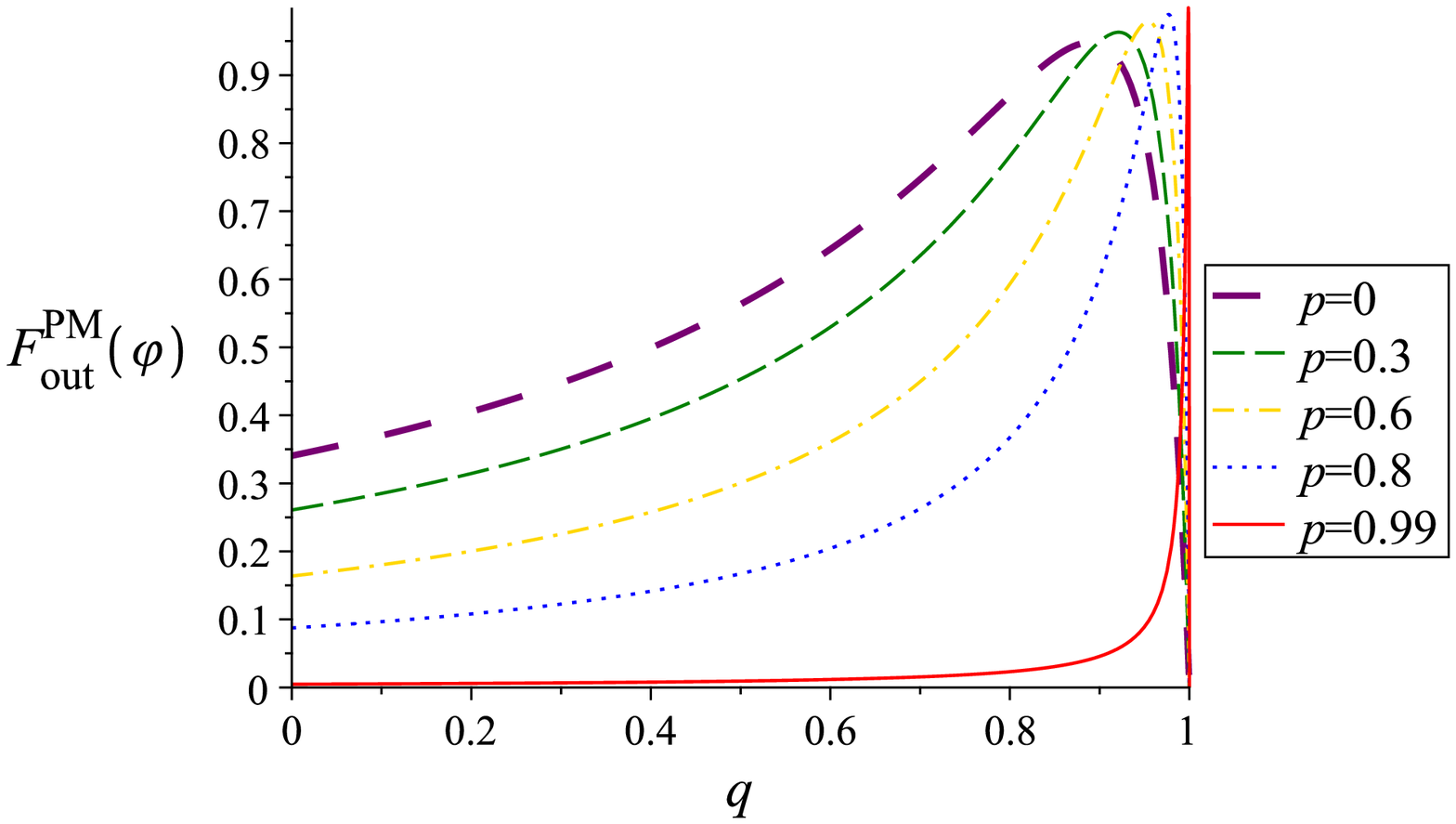}\label{qf22} }
	
	\caption{\small QFI of the single-qubit teleported state with respect to the phase parameter, $ \varphi $, as a function of (a) PM strength, $p$, fixing $\vartheta=\frac{3\pi}{4}$ and (b) PMR strength, $q$, fixing $\vartheta=\frac{\pi}{4} $; where we have chosen the acceleration parameter $r = 0.6$.}
	\label{qf2}   
\end{figure}   

\begin{figure}[ht!]
	\centering
	\subfigure[]{\includegraphics[width=5.7cm]{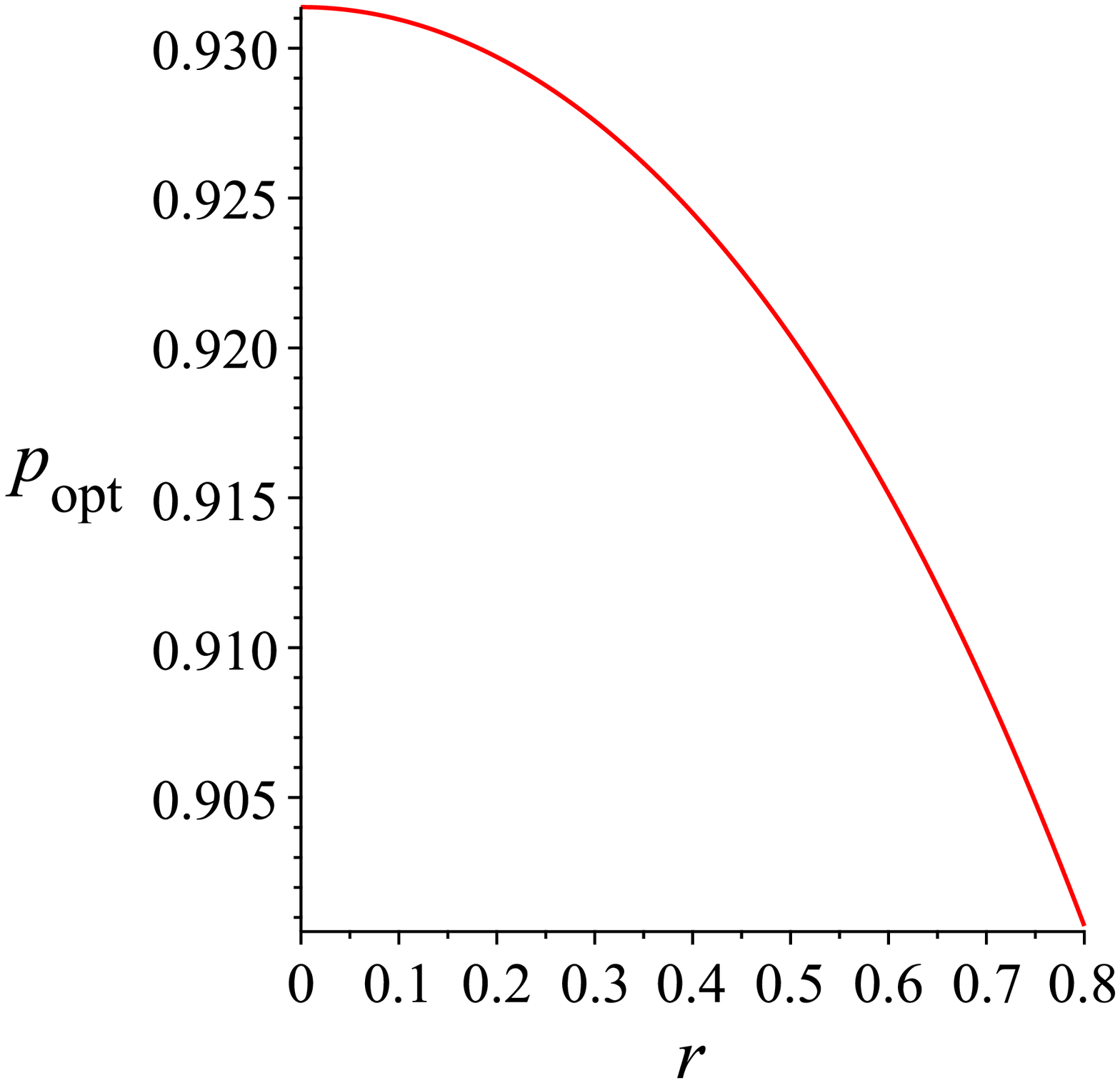}\label{opt1} }
	\subfigure[]{\includegraphics[width=5.7cm]{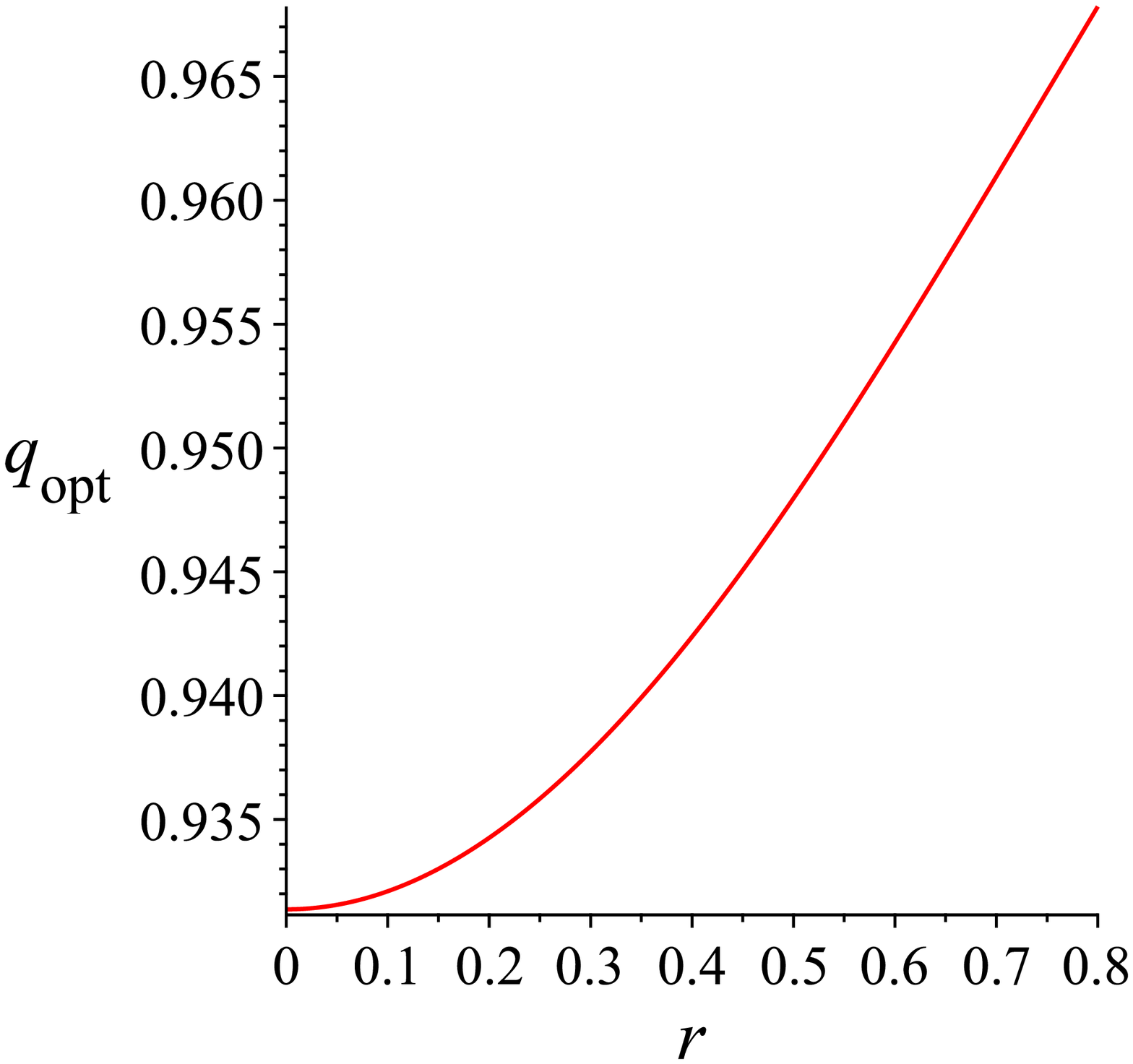}\label{opt2} }
	
	\caption{\small  The optimal value of PM and PMR strengths as  a function of acceleration parameter $ r $ for  (a) $q=0.6$ and $\vartheta=\frac{3\pi}{4}$ and (b) $p=0.6$ and $\vartheta=\frac{\pi}{4}$.}
	\label{opt}   
\end{figure}   

In Fig. \ref{qf2}, $ F_{\text{out}}^{\text{PM}}\left(\varphi \right) $ for single-qubit state teleportation, is plotted as  functions of PM as well as PMR strength for fixed value of acceleration parameter $r=0.6$ and the maximally entangled input state ($ \theta=\pi/2, \varphi=0  $). From Fig. \ref{qf21} it is seen that for $\frac{\pi}{2}\leq\vartheta<\pi $, with increase in PM strength $ F_{\text{out}}^{\text{PM}}\left(\varphi \right) $ increases to reach a maximum value and then it decreases with more increase of $ p $. Moreover, comparing the behavior of $ F_{\text{out}}^{\text{PM}}\left(\varphi \right) $ for different values of PMR strength, we see that with increase in $ q $, optimal estimation of the phase parameter occurs for larger values of $ p $. Nevertheless, increase of the PMR strength interestingly raises the optimal value of the QFI, leading to enhancement of the phase parameter estimation. We also see, in that range of $ \theta $, while for small values of $p$, the QFI may fall with an increase in $ q $, it can enhance as $ q $ increases for larger values of $p$. We obtain the same results investigating the behavior of $ F_{\text{out}}^{\text{PM}}\left(\varphi \right) $ versus $ q $ for $0<\vartheta\leq\frac{\pi}{2} $. In particular, in this range, the QFI may decrease with an increase in $ p $ for small values of $ q $, while it can exhibit increasing behavior as $ p $ increases for large values of $ q $ (see Fig.\ref{qf22}). 

\begin{figure}[ht!]
	\centering
	\subfigure[]{\includegraphics[width=6.6cm]{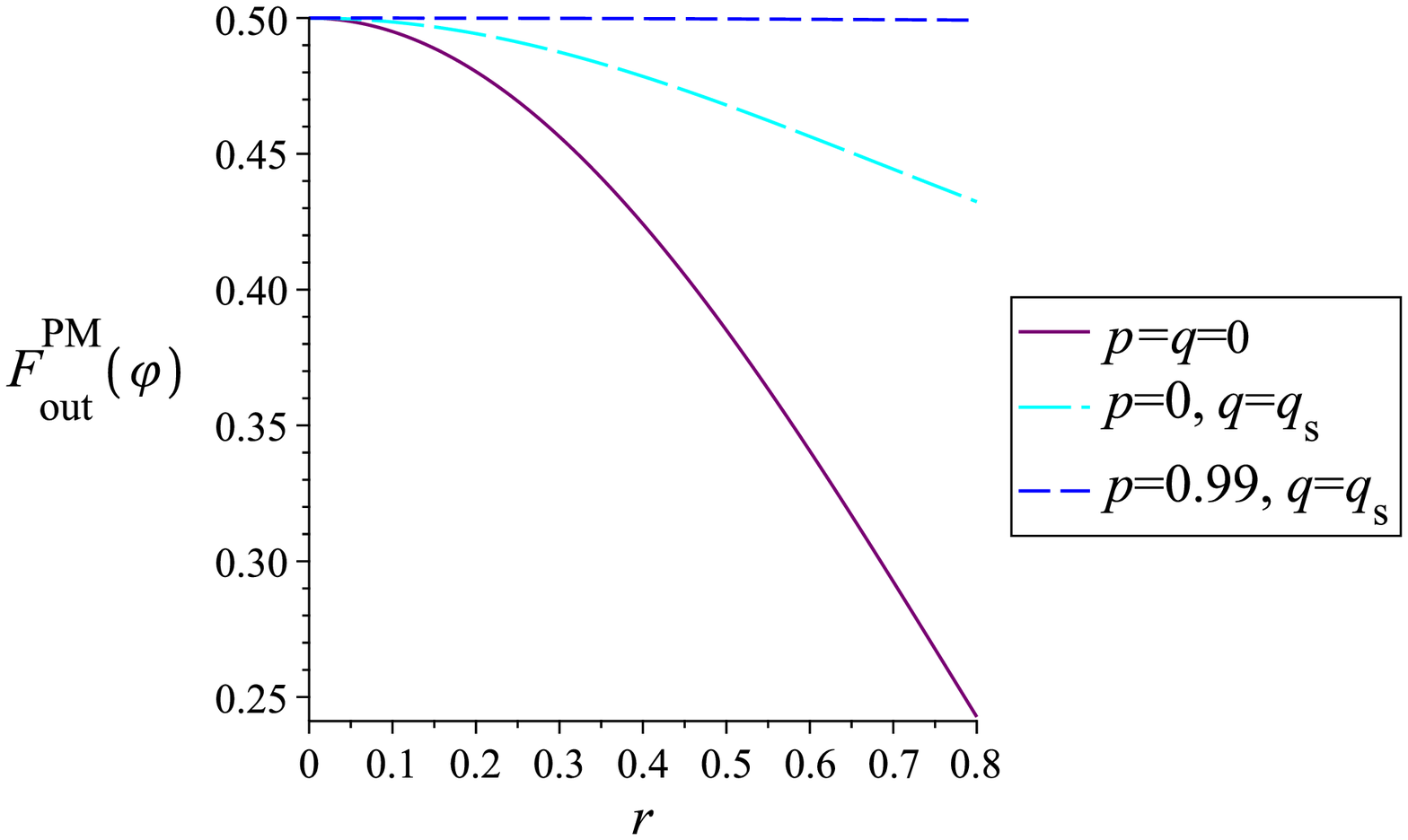}\label{ff1} }
	\subfigure[]{\includegraphics[width=6.6cm]{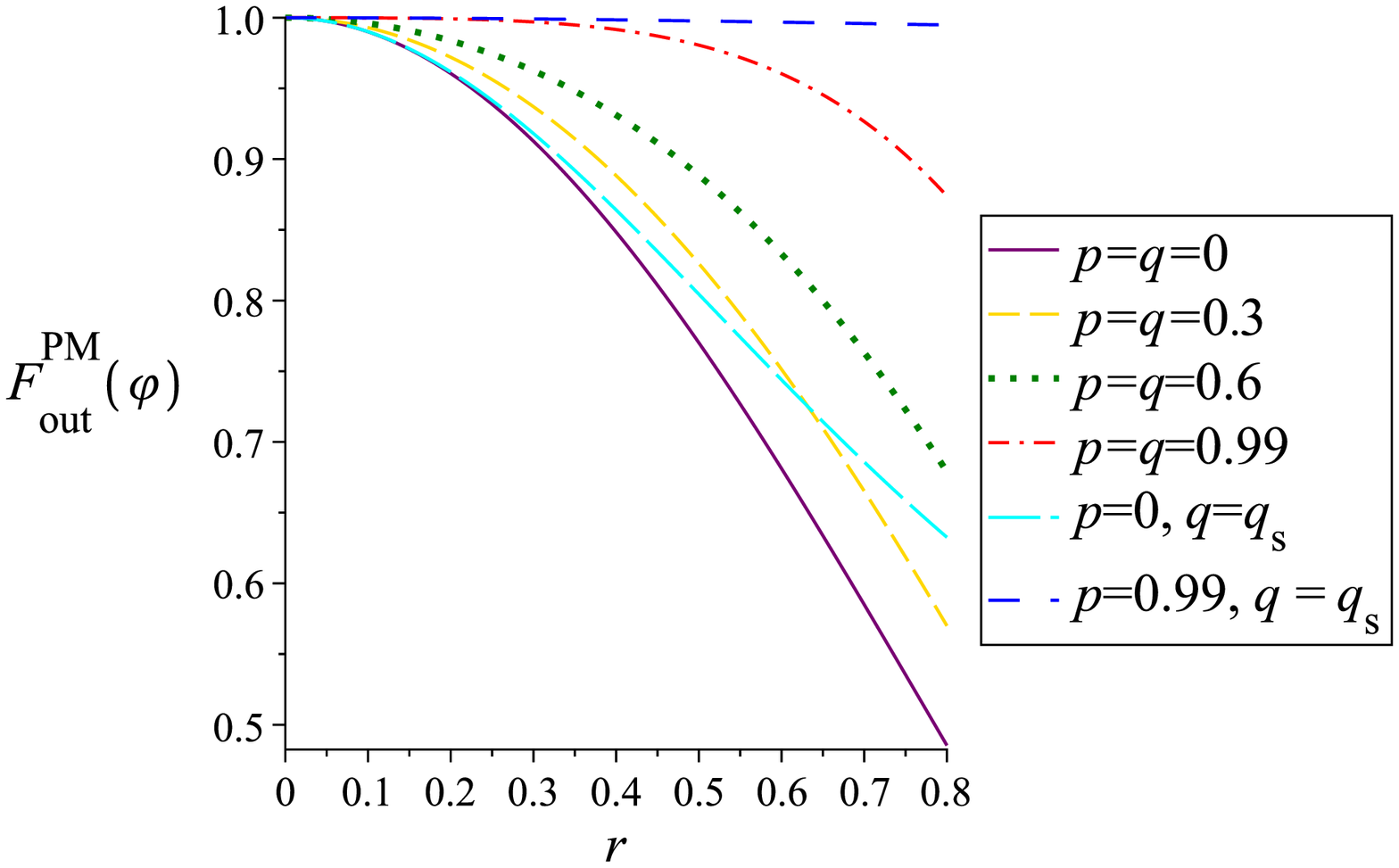}\label{ff2} }
	\subfigure[]{\includegraphics[width=6.6cm]{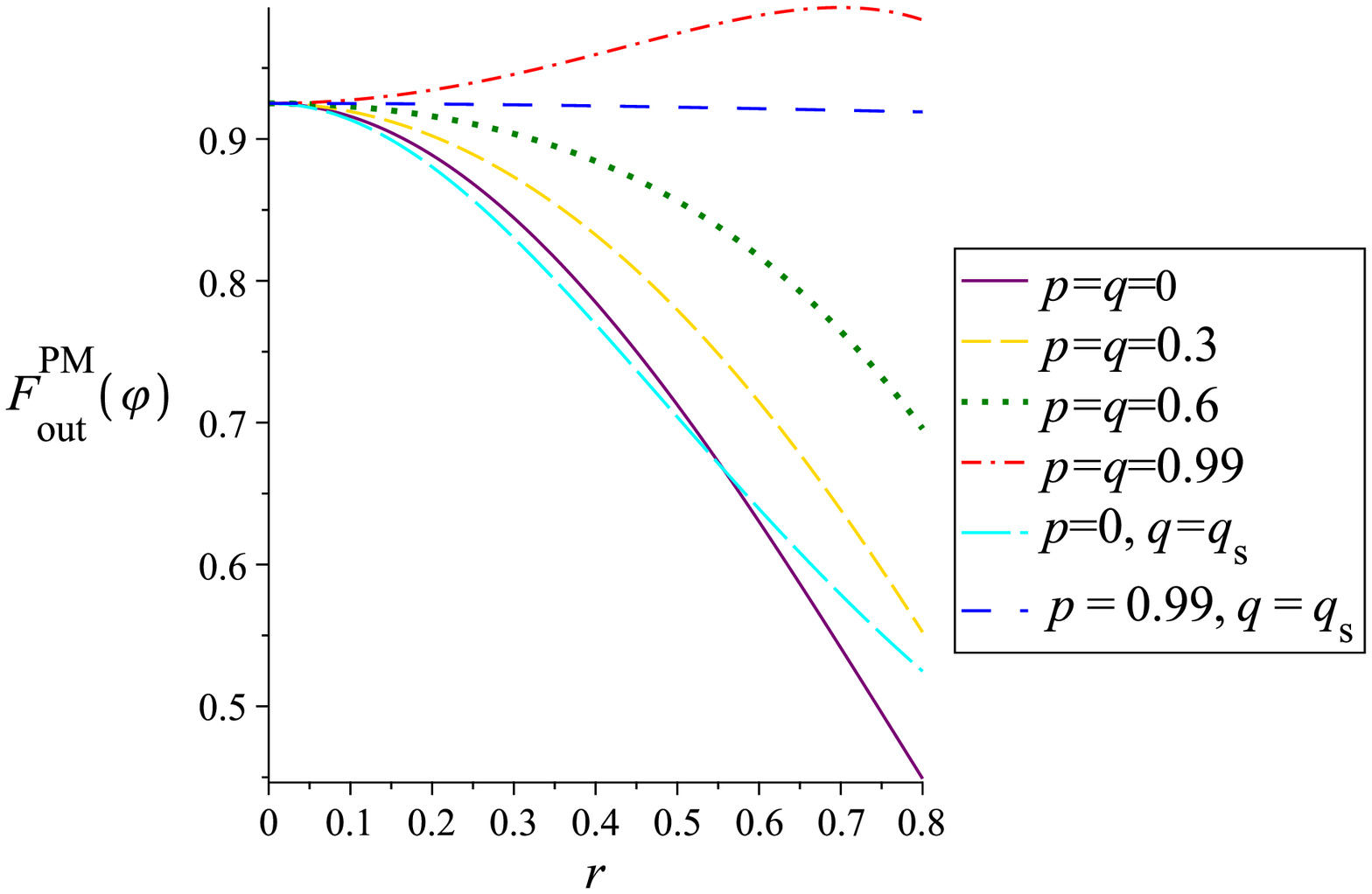}\label{ff3} }
	\caption{\small QFI of the single-qubit teleported state with respect to the phase parameter, $ \varphi $, as a function of acceleration parameter $ r $ by fixing $ \theta=\frac{\pi}{2} $  for (a) $ 0<\vartheta<\frac{\pi}{2} $, (b) $ \vartheta=\frac{\pi}{2} $ and (c) $\frac{\pi}{2}<\vartheta<\pi $.}
	\label{FF1}
\end{figure}

Considering the optimal behavior of QFI of the single-qubit teleported state with respect to the phase parameter as functions of $ p $ and $ q $, we obtain $ p_{\text{opt}} $ and $ q_{\text{opt}} $ as follows

\begin{eqnarray}\label{eq:opt}
p_{\text{opt}}=\frac{q\text{cos}^{2}r~\text{sin}^{2}\frac{\vartheta}{2}-\text{cos}\vartheta(1-q)}{\text{sin}^{2}\frac{\vartheta}{2}(1-q\text{sin}^{2}r)},  \nonumber \\
~~~~~~~~~~~~~~~~~q_{\text{opt}}=\frac{\text{sin}^{2}\frac{\vartheta}{2}(p+2(1-p)\text{cos}^{2}r)-1}{\text{sin}^{2}\frac{\vartheta}{2}(p+(1-p)\text{cos}^{2}r)-1}.
\end{eqnarray}
Figure \ref{opt} shows the variation of optimal values of $ p $ and $ q $, at which the the optimal estimation of the phase paramater happens, in terms of acceleration parameter $ r $.
We see that when the acceleration  increases, the optimal value of $p$,  $ p_\text{opt} $, decreases, i.e.,  the optimal estimation  is achieved by a weaker PM (see Fig. \ref{opt1}).
However, Fig. \ref{opt2} shows that a stronger PMR is required for attaining the  optimal value of $ F_{\text{out}}^{\text{PM}}\left(\varphi \right) $ when the accelerated observer moves with larger acceleration.

\par Behavior of the QFI with respect to phase parameter, $ F_{\text{out}}^{\text{PM}}\left(\varphi \right) $, as a function of acceleration parameter, $ r $, for different ranges of the channel parameter, $ \vartheta $, is investigated in Fig. \ref{FF1}. It is seen that for teleportation under pure Unruh channel (i.e., $ p=q=0 $), there is monotonous degradation in $ F_{\text{out}}^{\text{PM}}\left(\varphi \right) $ with increase in $r$. However, we find that the combined effect of PM and PMR with the same strength, ($ p=q $), leads to partially improvement of the the estimation precision of the phase parameter. Besides, when this common measurement strength increases $F_{\text{out}}^{\text{PM}}\left(\varphi \right) $ is protected much better for $\frac{\pi}{2}\leq\vartheta<\pi $; it even increases surprisingly with increase in acceleration for $\frac{\pi}{2}<\vartheta<\pi $, in the limit $ p\rightarrow1$ and $q\rightarrow1$. In addition, our numerical calculation shows that in order to protect the QFI with respect to $ \varphi $ and QRs of the teleported state against the Unruh effect, we can use the following \textit{special} choice for PMR strength \cite{X. Xiao}
\begin{equation}\label{a21}
q_{\text{s}}=1-\left(1-p\right)\text{cos}^{2}r.
\end{equation}
In fact, the Unruh noise may be approximately eliminated provided that the PM strength is sufficiently strong ($ p\rightarrow1 $) and the above choice for the PMR is applied (see blue dashed lines in Fig. \ref{FF1})

\par  Finally, if we intend to teleport only the information encoded into the phase parameter, we can manage the input state by choosing the weight parameter as $ \theta=\frac{\pi}{2} $, to estimate the phase parameter with the best precision; i.e., the best estimation of phase parameter is obtained if the input state is maximally entangled (see Eq. (\ref{EQ54})).

\par In the following, the effect of PM or PMR on QC of the teleported state of single-qubit are studied. Using the $l_{1}$-norm measure (Eq. (\ref{a7})), QC for the density matrix (\ref{a17}), can be obtained as follows
\begin{equation}\label{a19}
\mathcal{C}_{l_{1}}\left(\rho_{\text{out}}^{\text{PM}} \right)=|\frac{\text{sin}\vartheta\sqrt{\overline{pq}}\text{cos}~r\text{sin}~\theta}{N_{2}}|.
\end{equation}

In the case of teleportation without application of PM or PMR on the Unruh channel, i.e., $ p=q=0 $ and then $ N_{2}=1 $, we find
\begin{equation}\label{a20}
\mathcal{C}_{l_{1}}\left(\rho_{\text{out}} \right)=|\text{sin}\vartheta~\text{cos}~r\text{sin}~\theta|,
\end{equation}
which is QC of the teleported state under the pure Unruh decoherence. 

\par Investigating QC of the single-qubit teleported state as a function of $ r $ or studying its behavior versus PM and PMR strength for fixed value of the acceleration parameter, one can see that the results, qualitatively similar to $ F_{\text{out}}^{\text{PM}}\left(\varphi \right) $, are observed.

\par In order to determine the quality of teleportation, i.e., closeness of the teleported state to the input state, the fidelity \cite{Jozsa R} between $ \rho_{\text{in}} $ and $ \rho_{\text{out}} $ defined as $  f\left( \rho_{\text{in}},\rho_{\text{out}}\right)=\left\lbrace \text{Tr}\sqrt{\left(\rho_{\text{in}} \right)^{\frac{1}{2}}\rho_{\text{out}}\left( \rho_{\text{in}} \right)^{\frac{1}{2}} }\right\rbrace ^{2}=\langle\psi_{\text{in}}|\rho_{\text{out}}|\psi_{\text{in}}\rangle $, should be computed. Therefore, we obtain  
\begin{equation}\label{a23}
f=\frac{1}{N_{2}}\left[ \left(\frac{\mathcal{A}-\mathcal{D}}{2}+\mathcal{F}\text{cos}~2\varphi\right)\text{sin}^{2}\theta  +\mathcal{D}\right], 
\end{equation}

\begin{figure}[ht!]
	\centering
	\subfigure[]{\includegraphics[width=6cm]{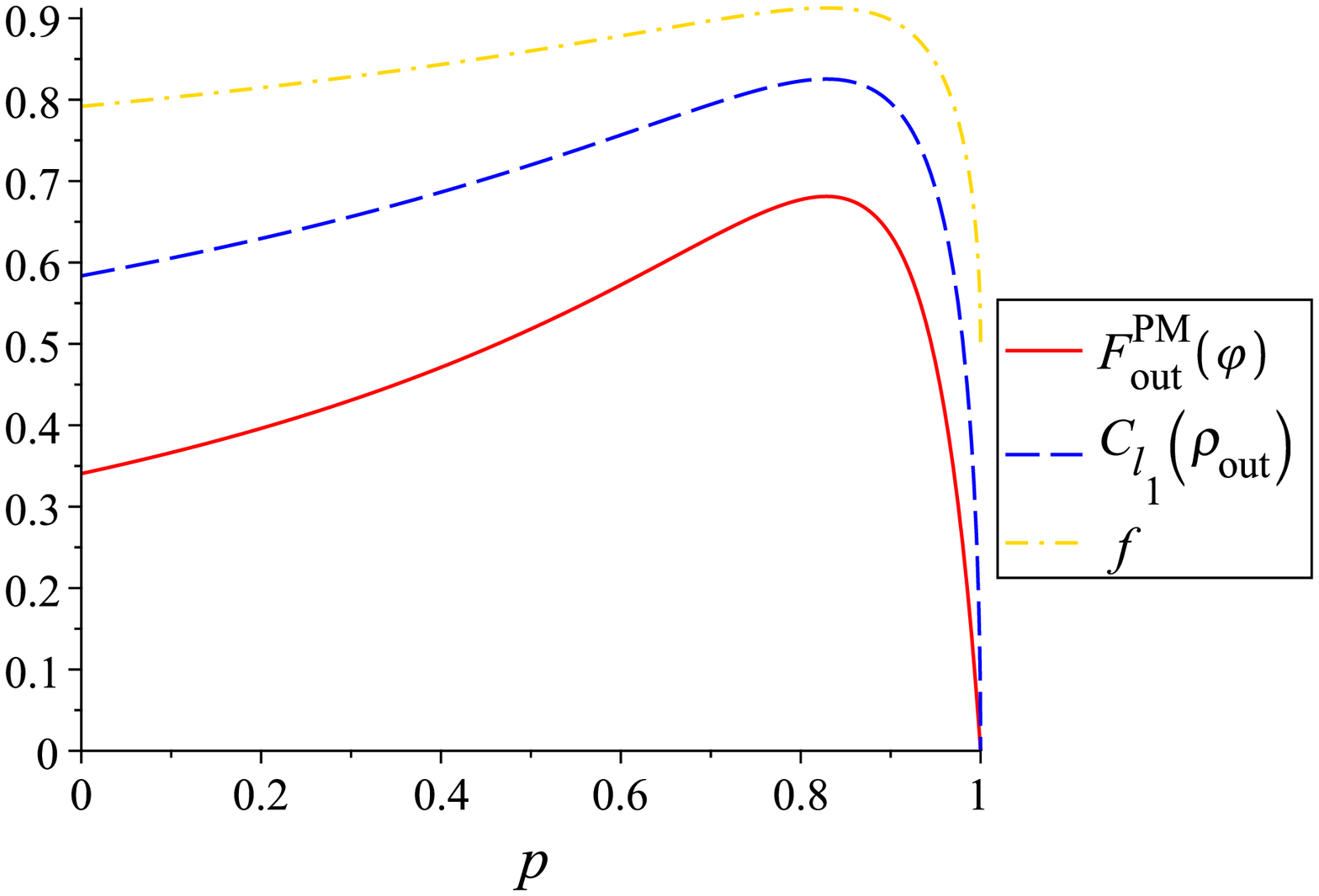}\label{sim1} }
	\subfigure[]{\includegraphics[width=6cm]{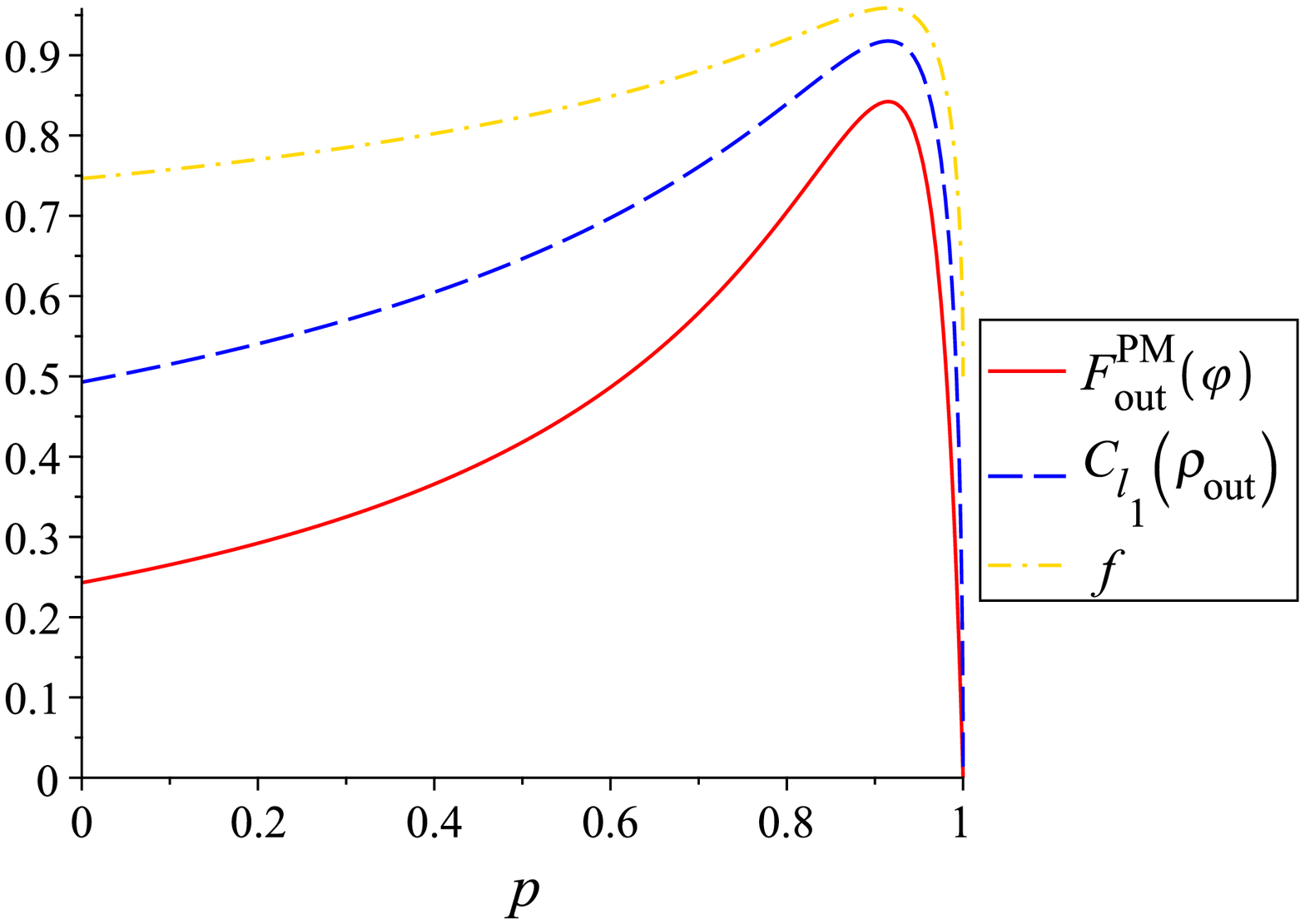}\label{sim2} }
	\subfigure[]{\includegraphics[width=6cm]{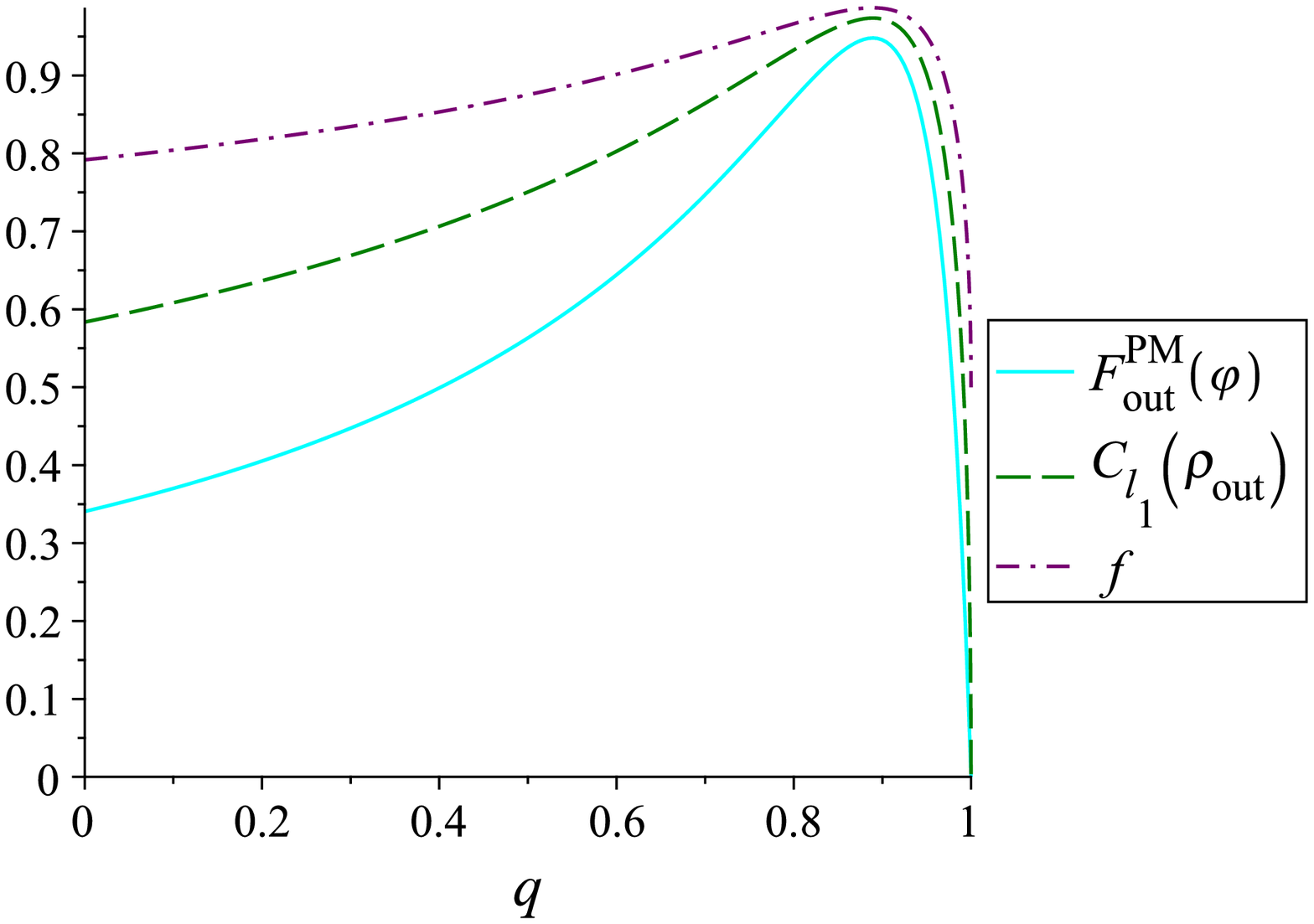}\label{sim3} }
	\subfigure[]{\includegraphics[width=6cm]{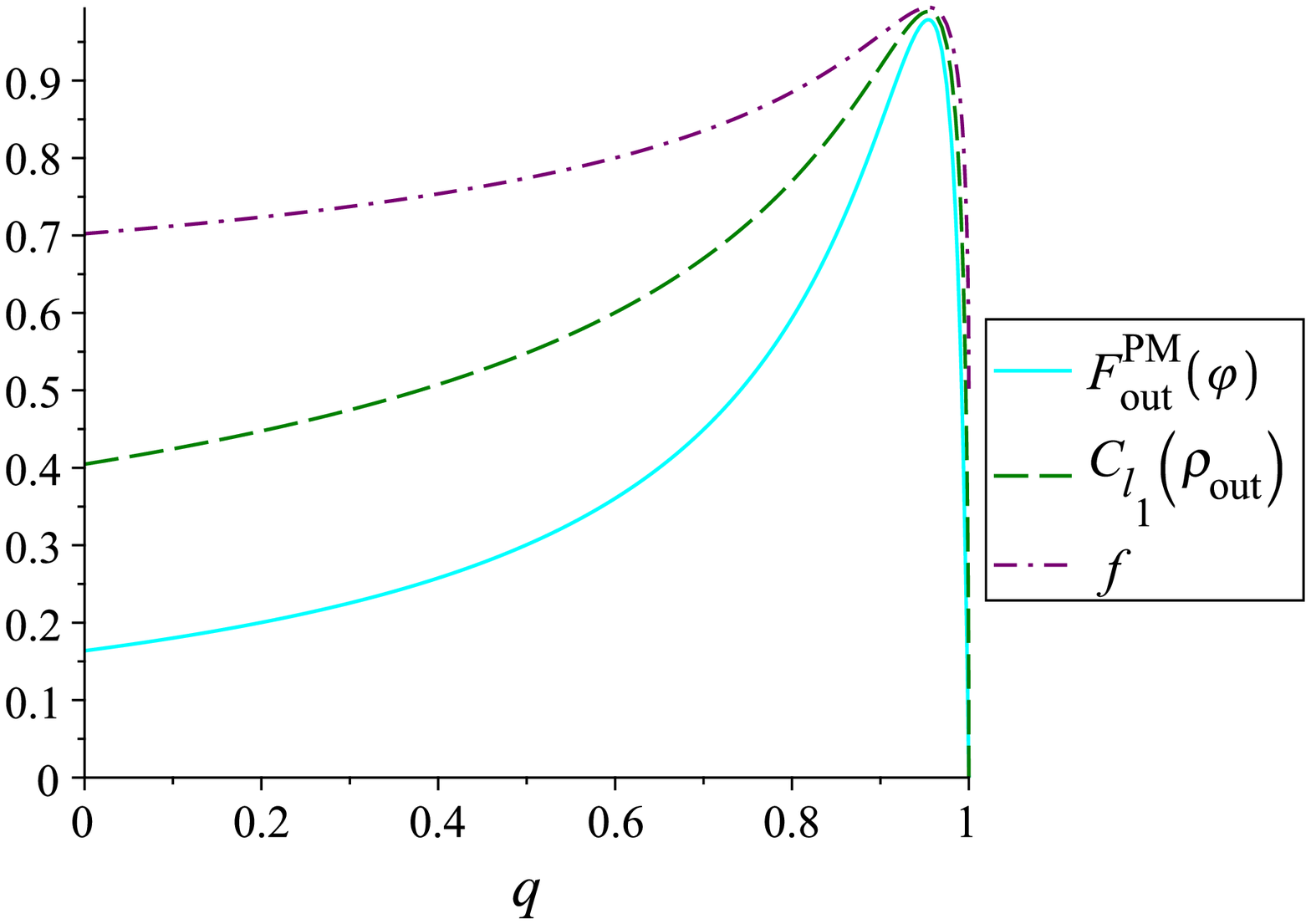}\label{sim4} }
	\caption{\small Comparing $ F_{\text{out}}^{\text{PM}}\left(\varphi \right) $ and QC of teleported state as well as teleportation fidelity in single-qubit scenario 
	by fixing $ \theta=\frac{\pi}{2},\varphi=0 $, (a) versus $p$ in the absence of PMR (i.e., $ q=0 $), (b) versus $p$ in the presence of PMR with $ q=0.6 $, (c) versus $q$ in the absence of PM (i.e., $ p=0 $), and (d) versus $q$ in the presence of PM with $ p=0.6 $. }
	\label{sim}   
\end{figure} 

\par Examining the behavior of the teleportation fidelity, we  again see that the results obtained for fidelity is the same as the obtained results for QC and $ F_{\text{out}}^{\text{PM}}\left(\varphi \right) $, i.e., fidelity degrades with increase in $ r $ under pure Unruh effect. However, the combined effects of PM and PMR for channel parameter lying in the region $\frac{\pi}{2}\leq\vartheta<\pi $, can improve the quality of teleportation and it may even enhance with increase in acceleration for $\frac{\pi}{2}<\vartheta<\pi $ in the limit $ p,q\rightarrow1 $. Moreover, Unruh decoherence is approximately eliminated for all values of $ \vartheta  $, with $ q = q_{opt} $ and in the limit $ p\rightarrow1 $, consequently the teleportation process may be implemented with better quality.  
On the other hand,   investigating the fidelity of the single-qubit teleportation as functions of PM as well as PMR strength, we find that, similar to $ F_{\text{out}}^{\text{PM}}\left(\varphi \right) $ and QC, with proper selection of the channel parameter $ \vartheta $ the quality of teleportation may be enhanced with increase in $ p $ or $ q $ to reach a maximum value. Besides, in the range $0<\vartheta\leq\frac{\pi}{2}$ ($\frac{\pi}{2}<\vartheta<\pi $), analyzing the fidelity behavior as a function of $q$ ($ p $), we see that  it may be decreased (improved) with an increase in $ p $ ($ q $) for small values of $ q $ ($ p $),  while it can exhibit increasing behavior  as $ p $ ($ q $) increases for large values of q($ p $). In addition, optimal teleportation fidelity becomes greater with increase in $ q $ or $ q $, hence the teleportation process is done more successfully.   

\par Finally, in Figs. \ref{sim1} and \ref{sim2} we compare and illustrate the harmonic behavior of $ F_{\text{out}}^{\text{PM}}\left(\varphi \right) $, QC and teleportation fidelity as functions of PM strength for $\frac{\pi}{2}<\vartheta<\pi$, and PMR strength for $0<\vartheta\leq\frac{\pi}{2}$, in the case of single-qubit teleportation. We conclude that for both $q=0$ and $q\ne0$, the PM strength which optimizes the estimation precision of the phase parameter, is the strength at which the quality of teleportation is the best and the coherence of the output single-qubit state reaches to its maximum value. Investigating the behavior of the above mentioned quantities as functions of PMR strength, we achieve the same results (see Figs. \ref{sim3} and \ref{sim4}).

\section{Two-qubit teleportation under the Unruh noise channel\label{2qbtele}}
In order to study the QRs and QFIs of the teleported two-qubit   state through the Unruh channel, $ |\psi_{in}\rangle=\text{cos}~\theta/2 |10\rangle+e^{i\varphi}~\text{sin}~\theta/2 |01\rangle,\ 0\leq\theta\leq\pi,\ 0\leq\varphi\leq2\pi  $ is considered as the input state in the teleportation process. We follow Kim and Lee's two-qubit teleportation protocol \cite{Lee J}, and use two copies of  shared state  (\ref{a16}) between Alice and Rob as the quantum channel, i.e. again $ \rho_{A,I}=\rho_{ch} $. Using Eq. (\ref{a3}), we obtain the output state as
\begin{equation}\label{b1}
\rho_{\text{out}}^{\text{PM}}=\frac{1}{N_{2}^{2}}\left(\begin{array}{cccc}
\mathcal{A}\mathcal{D} & 0& 0 & 0 \\
0 & \mathcal{A}^{2}\text{sin}^{2}\frac{\theta}{2}+\mathcal{D}^{2}\text{cos}^{2}\frac{\theta}{2}&2\mathcal{F}^{2}e^{i\varphi}\text{sin}~\theta & 0 \\
0 & 2\mathcal{F}^{2}e^{-i\varphi}\text{sin}~\theta&\mathcal{A}^{2}\text{cos}^{2}\frac{\theta}{2}+\mathcal{D}^{2}\text{sin}^{2}\frac{\theta}{2} &0 \\
0 & 0 & 0& \mathcal{A}\mathcal{D}  \\
\end{array}\right),
\end{equation}
where $ \mathcal{A} $, $ \mathcal{D} $ and $ \mathcal{F} $ are determined by Eq. (\ref{a18}).

We study how  the  PM and  PMR help us to enhance the  QRs and QFIs of the teleported state in the presence of the Unruh effect. Using density matrix (\ref{b1}), the corresponding quantum coherence  is obtained as follows 
\begin{equation}
\mathcal{C}_{l_{1}}\left(\rho_{\text{out}}^{\text{PM}} \right)=4|\frac{\mathcal{F}^{2}\text{sin}~\theta}{N_{2}^{2}}|,
\end{equation}
\par The results, obtained for QC of the teleported two-qubit state under Unruh noise channel, are similar to single-qubit teleportation.

\begin{figure}[ht!]
	\centering
	\subfigure[]{\includegraphics[width=6.6cm]{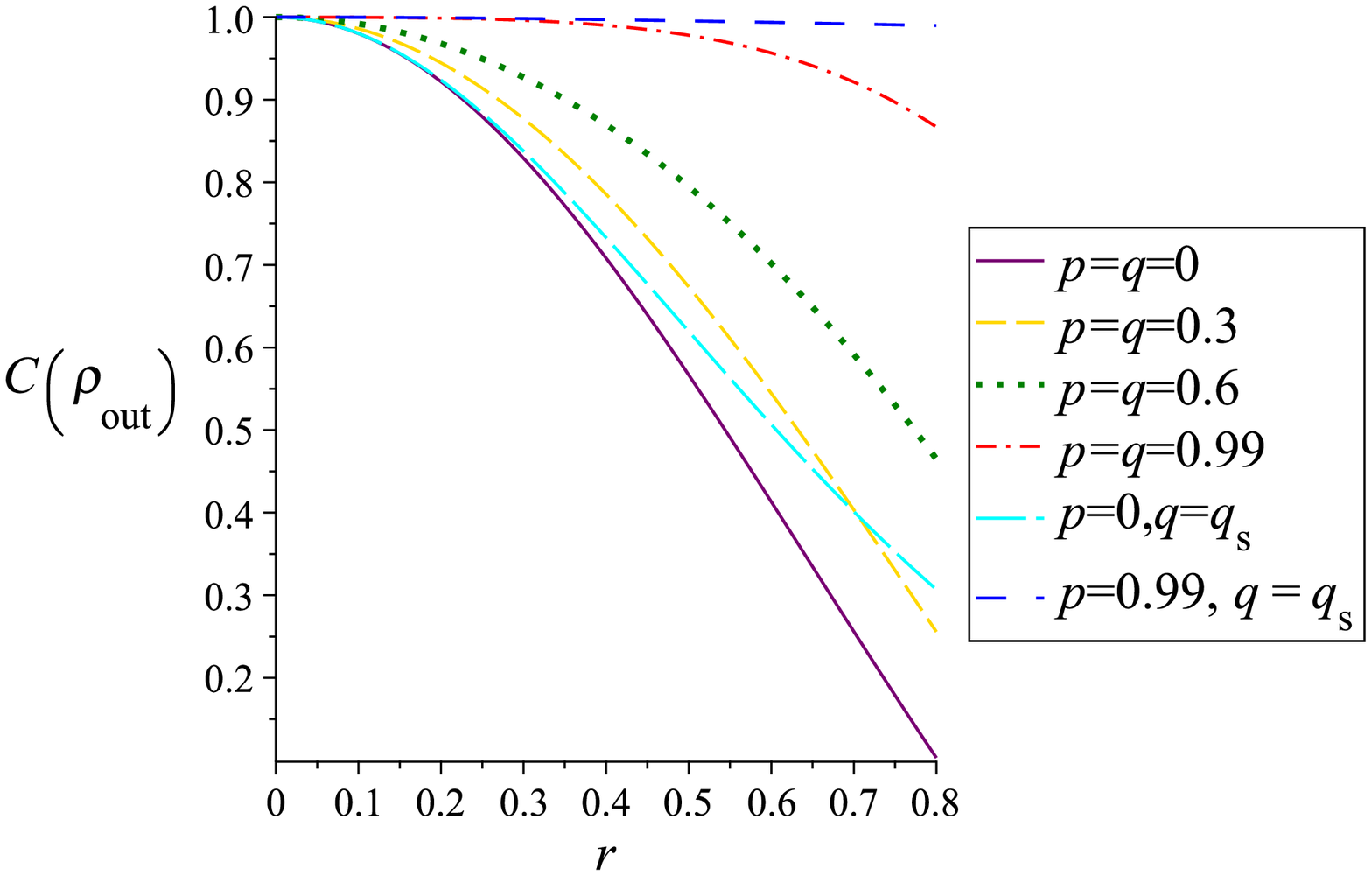}\label{F151} }
	\subfigure[]{\includegraphics[width=6.6cm]{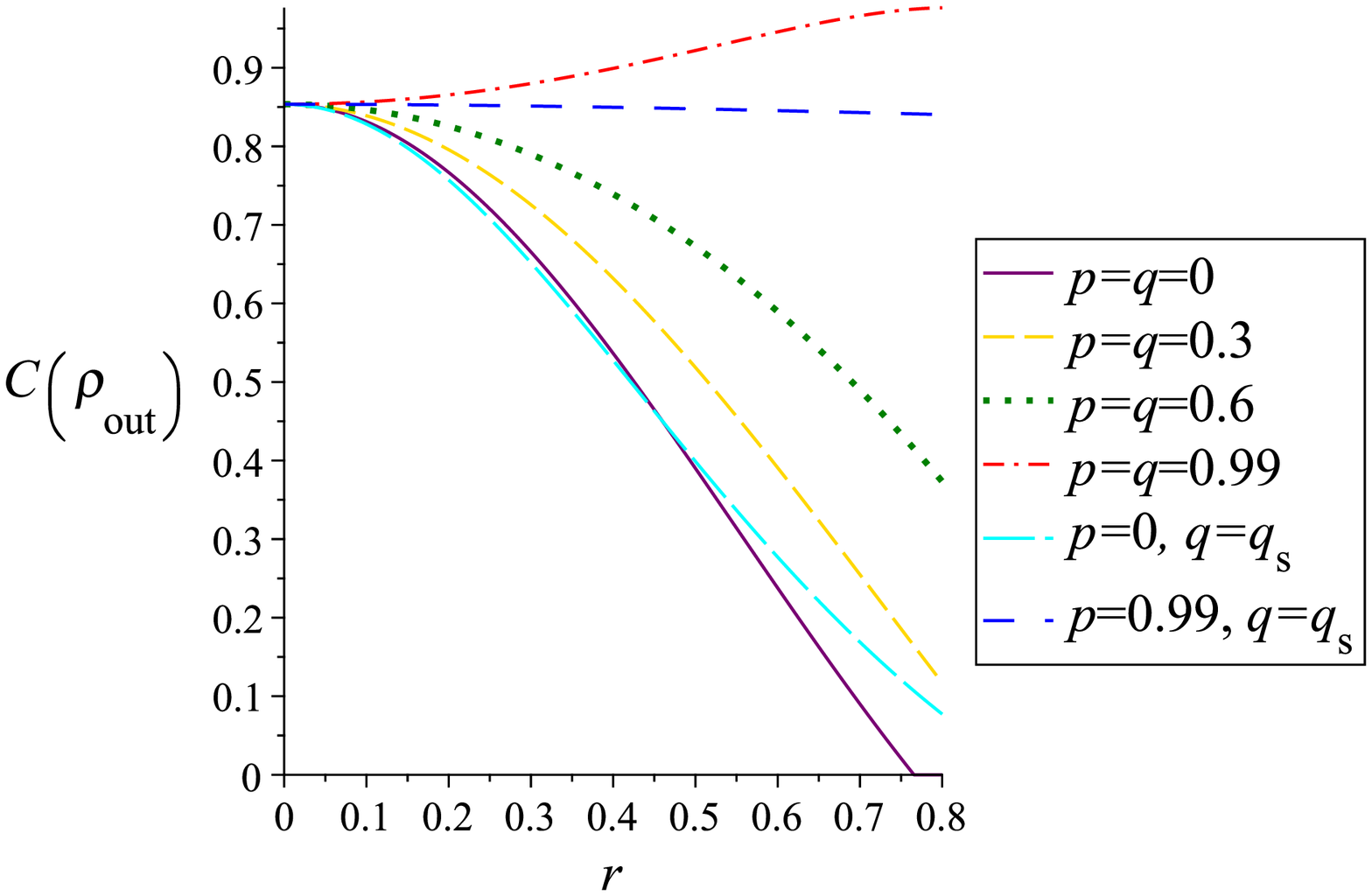}\label{F152} }
	\caption{\small Entanglement of the teleported two-qubit state as a function of acceleration parameter $ r $ by fixing $ \theta=\frac{\pi}{2} $ for (a) $ 0<\vartheta\leq\frac{\pi}{2} $, and (b) $\frac{\pi}{2}<\vartheta<\pi $.}
	\label{F15}   
\end{figure}
\par According to the Eqs. (\ref{a8}) and (\ref{b1}), entanglement of the teleported two-qubit state is obtained as
\begin{equation}
C\left(\rho_{\text{out}}^{\text{PM}} \right)=2Max\left\lbrace 0,2|\frac{\mathcal{F}^{2}\text{sin}~\theta}{N_{2}^{2}}|-|\frac{\mathcal{A} \mathcal{D}}{N_{2}^{2}}|\right\rbrace.
\end{equation}

\par In Fig. \ref{F15}, we plot the concurrence of the teleported two-qubit state as a function of acceleration parameter $ r $ for different strengths of PM and PMR. It is clear that the entanglement absolutely decreases with increase in acceleration under the pure Unruh decoherence. However, it can be amplified with combined action of PM and PMR for all values of initial channel parameter $ \vartheta $. In fact, when the strength of PM increases, the entanglement degradation decreases, especially in the limit $ p=q\rightarrow1 $, entanglement is approximately protected against the Unruh decoherence. Surprisingly, as seen in Fig. \ref{F15}, in that limit, entanglement of the teleported state may increase under the Unruh effect for initial channel parameter lying in the region $\frac{\pi}{2}<\vartheta<\pi $. In addition, it is seen that the entanglement is also improved by applying $ q_{s} $ even without first PM (i.e., $ p=0 $) for $ 0<\vartheta\leq\frac{\pi}{2} $.
\begin{figure}[ht!]
	\centering
	\subfigure[]{\includegraphics[width=6.6cm]{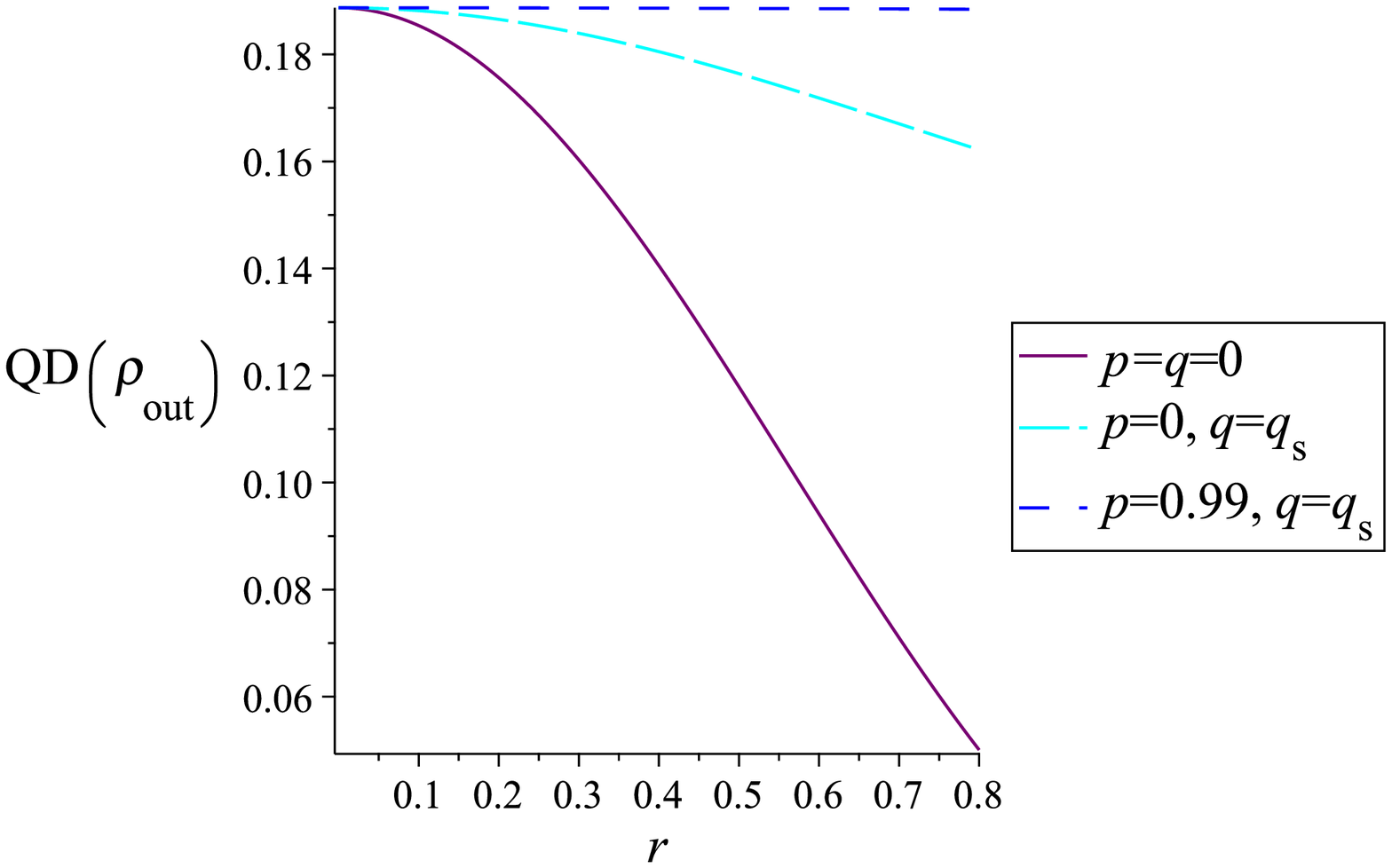}\label{F171} }
	\subfigure[]{\includegraphics[width=6.6cm]{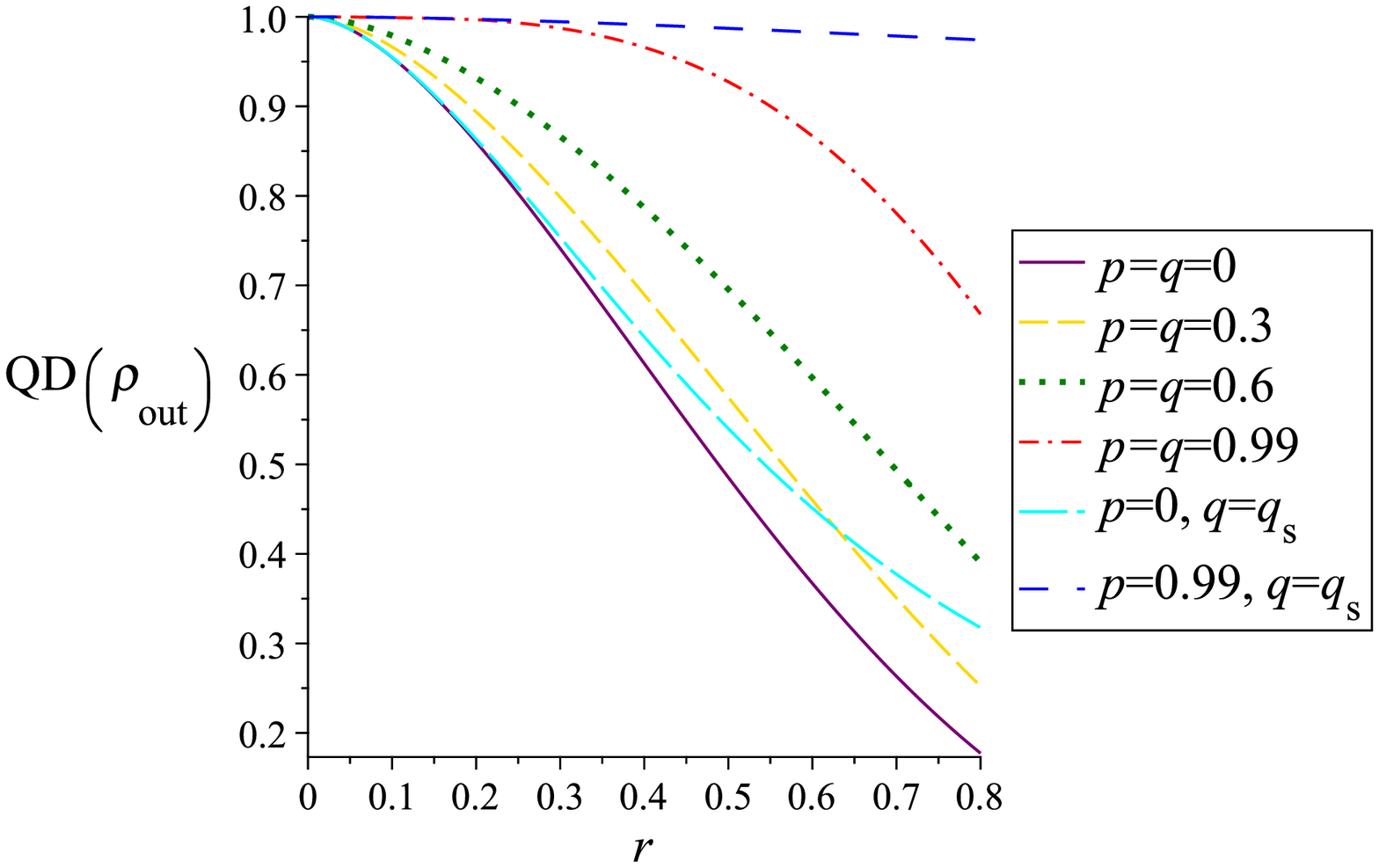}\label{F172} }
	\subfigure[]{\includegraphics[width=6.6cm]{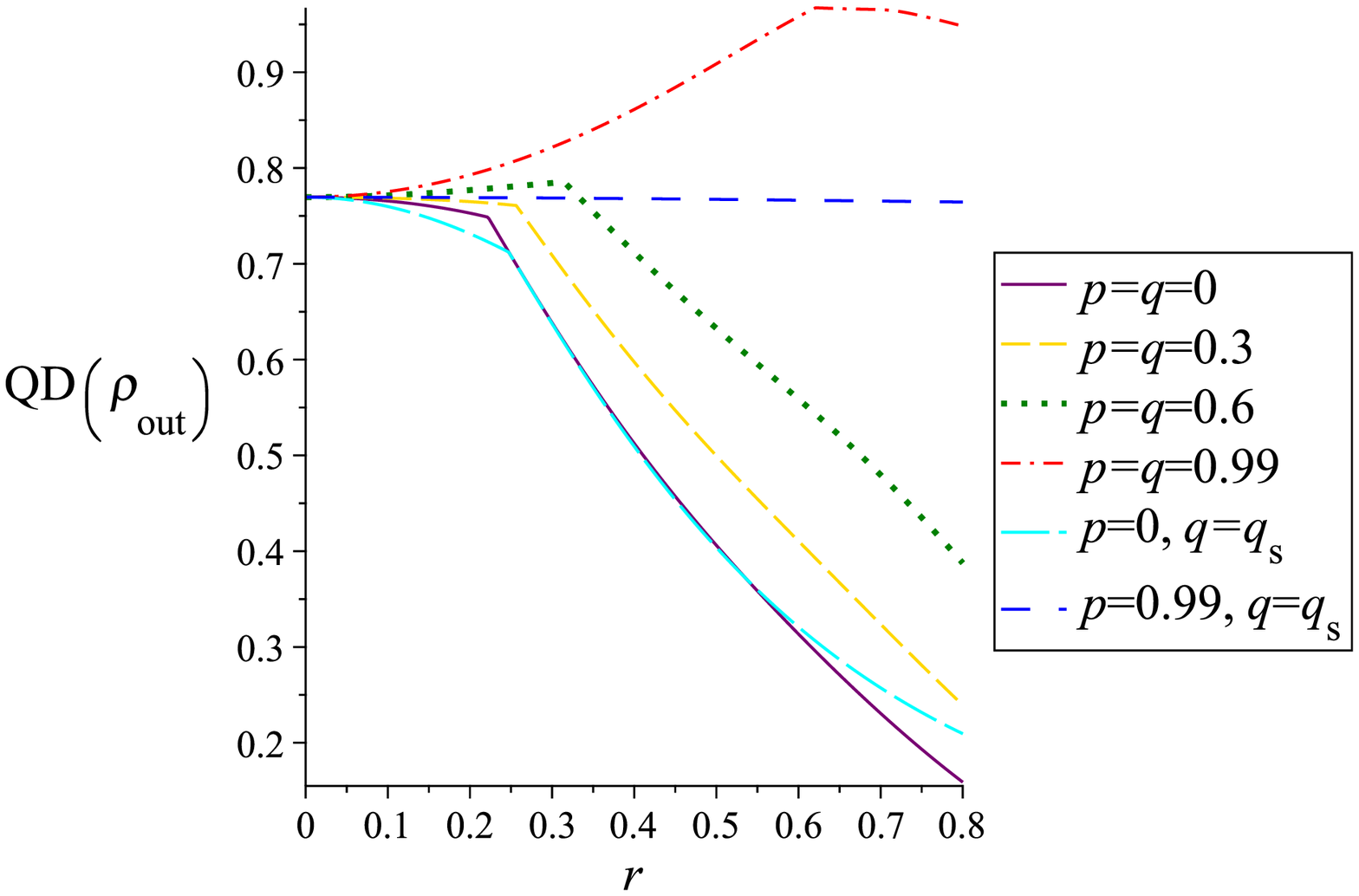}\label{F173} }
	\caption{\small QD of the teleported two-qubit state as a function of acceleration parameter $ r $ by fixing $ \theta=\frac{\pi}{2} $ and $ \varphi=0 $ for (a) $ 0<\vartheta<\frac{\pi}{2} $, (b) $\vartheta=\frac{\pi}{2} $, and (c) $\frac{\pi}{2}<\vartheta<\pi $}
	\label{F17}   
\end{figure}
\par Considering the behavior of QD as a function of acceleration parameter $ r $ for different PMs strength, we see that combined action of PM and PMR can raise QD for $ \vartheta $ lying in the range $\frac{\pi}{2}\leq\vartheta<\pi $ (see Fig.\ref{F17}). In particular, in the limit $ p,q\rightarrow1 $, QD may increase with applying PMs for $\frac{\pi}{2}<\vartheta<\pi $. Moreover, if we choose $ q=q_{opt} $, QD can increase even in the absence of first PM (i.e., $ p=0 $) for $ 0<\vartheta\leq\frac{\pi}{2} $. 
 
Next we consider the QFI of the teleported two-qubit state. In order to calculate the QFIs associated with the  weight and phase parameters encoded into the quantum state (\ref{b1}), we follow the method proposed in \cite{LJZ+14,J. Liu11} to block diagonal states, because state (\ref{b1}) is X-type and hence it becomes block diagonal after changing the order of the basis vectors. A block diagonal state can be written as $ \rho=\bigoplus^{n}_{i=1}\rho_{i}$, where $ \bigoplus $ represents the direct sum. The SLD operator is block diagonal here, i.e. $L=\bigoplus^{n}_{i=1}L_{i}$, where $L_{i}$ indicates the corresponding SLD operator for $\rho_{i}$. 
	For 2-dimensional blocks, it can be proved that the SLD operator for the $i$th block is given by~\cite{J. Liu11} 
\begin{equation}
L_{i}=\frac{1}{\mu_{i}}\left[\partial_{x}\rho_{i}+\xi_{i}\rho^{-1}_{i} -\partial_{x}\mu_{i} \right], 
\end{equation}
where $\xi_{i}=2\mu_{i}\partial_{x}\mu_{i}-\partial_{x}P_{i}/4$, in which $\mu_{i}=\text{Tr}\rho_{i}/2$ and $P_{i}=\text{Tr}\rho^{2}_{i}$. If det$ \rho_{i}=0 $, $\xi_{i}$ vanishes. 

Using the above mentioned method, we find the QFIs of the two-qubit teleported state with respect to weight and phase parameters as follows 
\begin{equation}
F_{\text{out}}^{\text{PM}}\left(\theta \right)=\frac{1}{N_{2}^{2}}\left[ \zeta+\frac{8\mathcal{A}^{2}\mathcal{D}^{2}\left( \zeta^{2}-16\mathcal{F}^{4}\right) }{\zeta\left[\left(\left(\mathcal{A}^{2}-\mathcal{D}^{2} \right)^{2} -16\mathcal{F}^{4} \right)\text{cos}~2\theta -\left(\zeta^{2}+4\left(\mathcal{A}^{2}\mathcal{D}^{2}-4\mathcal{F}^{4} \right)  \right)  \right] }\right], 
\end{equation}
\begin{equation}
F_{\text{out}}^{\text{PM}}\left(\varphi \right)= \frac{16\mathcal{F}^{4}\text{sin}^{2}\theta}{\zeta N_{2}^{2}},
\end{equation}
where $ \zeta=\mathcal{A}^{2}+\mathcal{D}^{2}  $. Surprisingly, investigating QFI of teleported two-qubit state under Unruh channel, we obtain the results similar to single-qubit teleportation.
\begin{figure}[ht!]
	\centering
	\subfigure[]{\includegraphics[width=6.6cm]{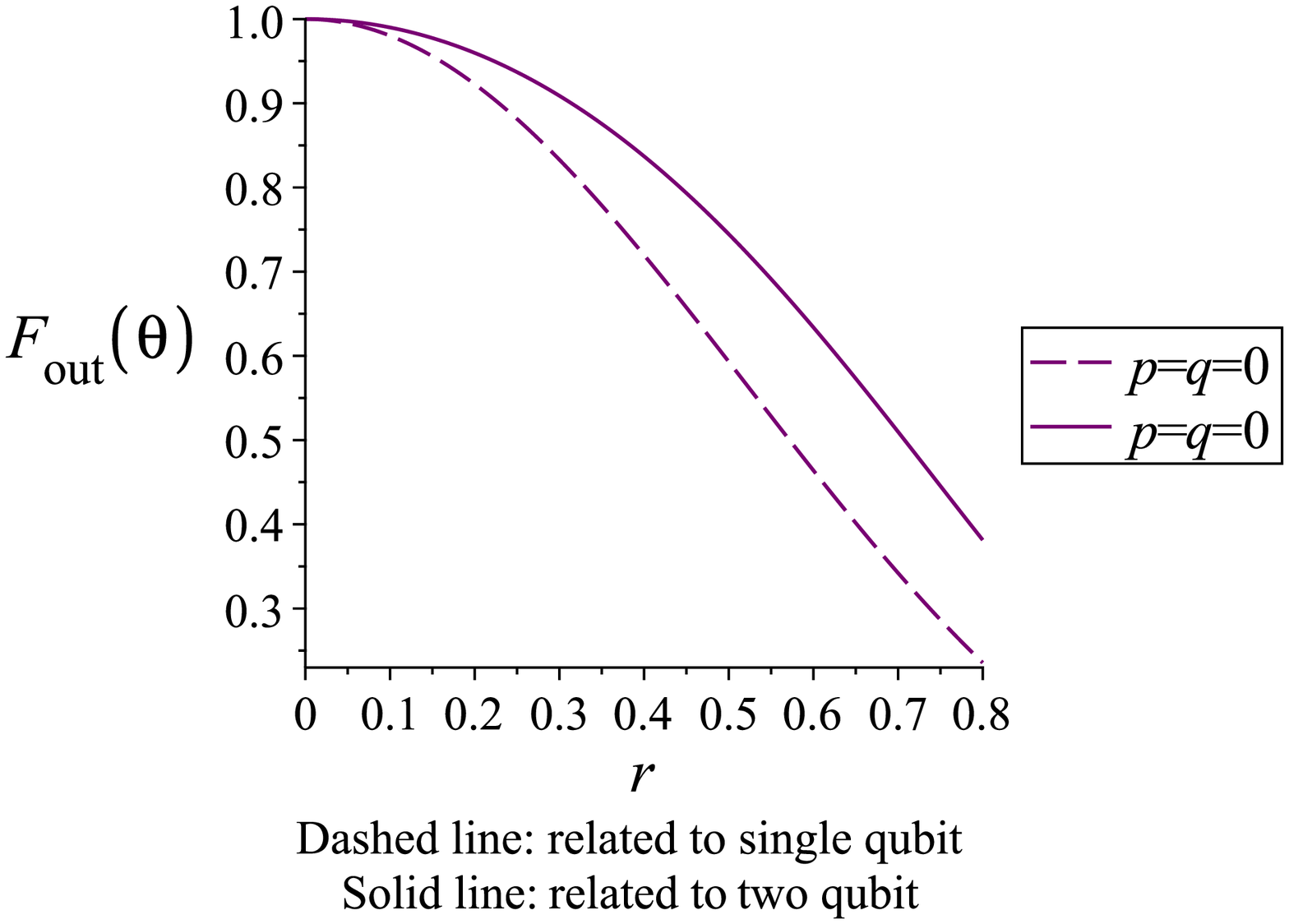}\label{F131} }
	\subfigure[]{\includegraphics[width=6.6cm]{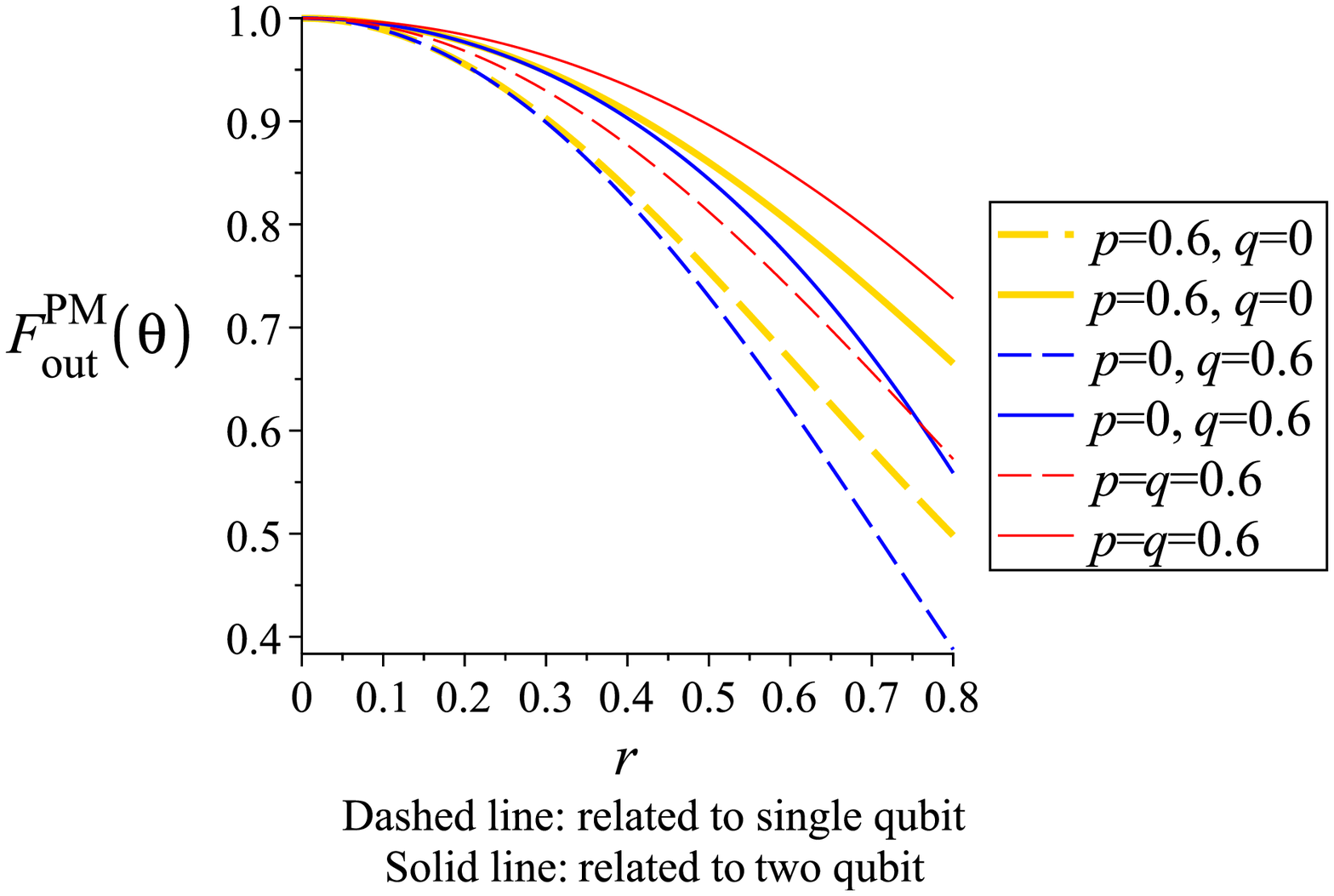}\label{F132} }
	\caption{\small Comparing $ F_{\text{out}}^{\text{PM}}\left(\theta \right) $ of the teleported single- and two-qubit states, fixing $ \theta=\frac{\pi}{2} $ and $ \vartheta=\frac{\pi}{2}$ (a) in the absence of measurements, (b) in the presence of measurements.}
	\label{F13}   
\end{figure} 
\par
\begin{figure}[ht!]
	\centering
	\subfigure[]{\includegraphics[width=6.6cm]{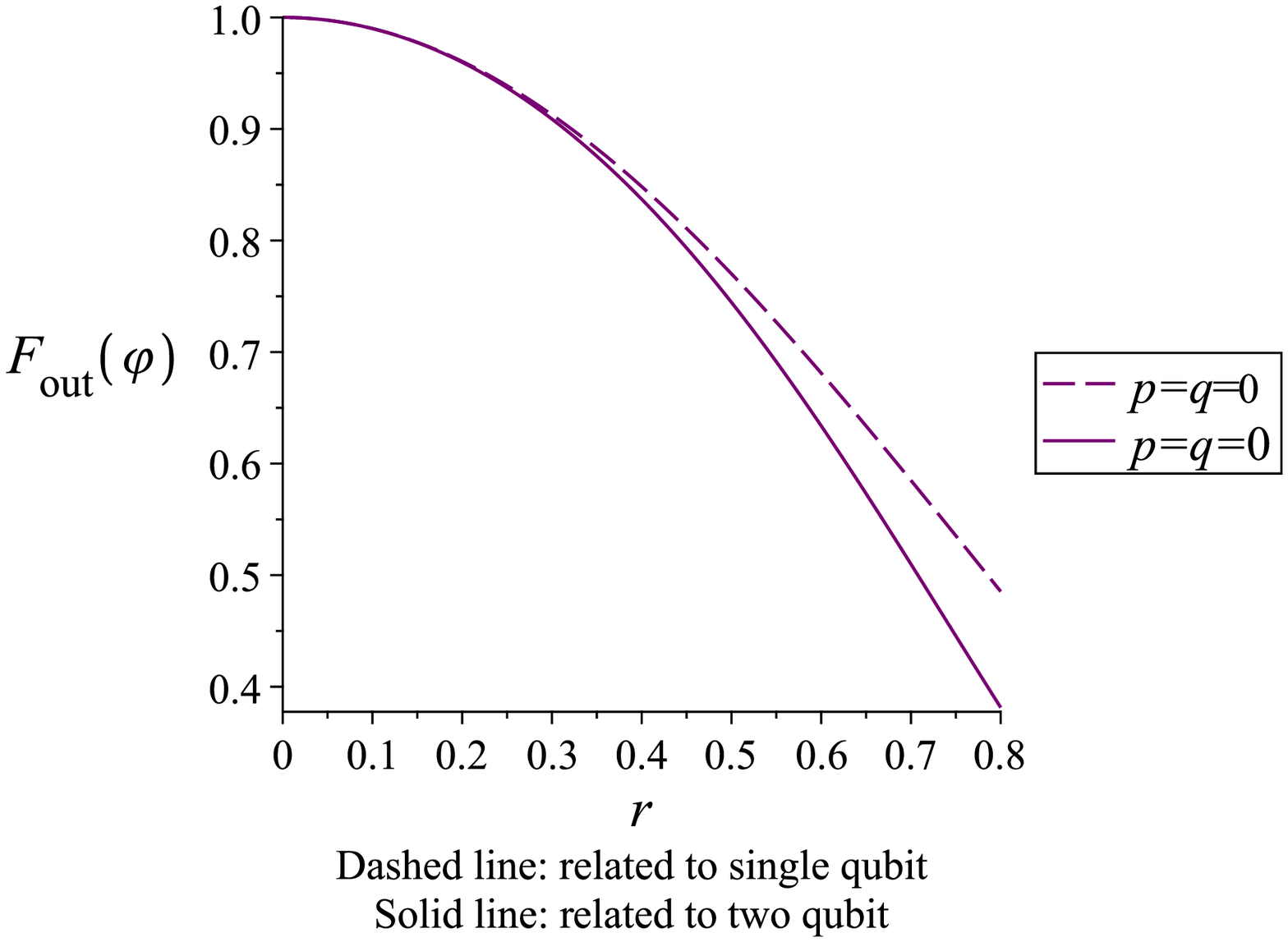}\label{F111} }
	\subfigure[]{\includegraphics[width=6.6cm]{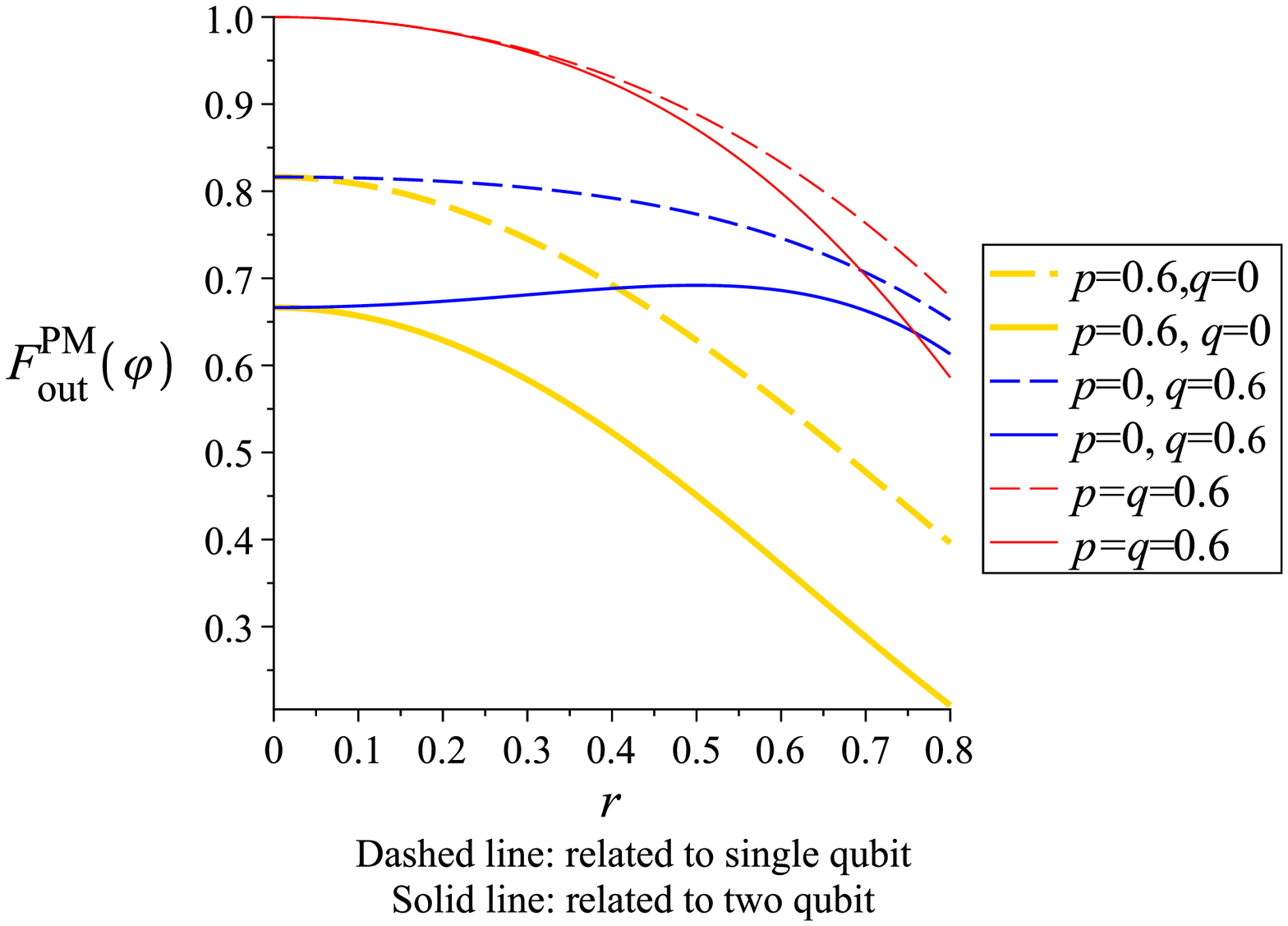}\label{F112} }
	\caption{\small Comparing $ F_{\text{out}}^{\text{PM}}\left(\varphi \right) $ of the teleported single- and two-qubit states 
		fixing $ \theta=\frac{\pi}{2} $ and $ \vartheta=\frac{\pi}{2}$ (a) in the absence of measurements, (b) in the presence of measurements.}
	\label{F11}   
\end{figure}

  In Figs. \ref{F13} and \ref{F11}, we compare the QFI of both single- and two-qubit teleported states (supposing that $ \theta $ or $ \varphi $ carries the same information in both cases). In Fig. \ref{F13}, we see that the information encoded in the weight parameter $ \theta $ is better protected against Unruh effect during teleportation of two-qubit state, comparing it with the single-qubit scenario. Nevertheless, extraction of information encoded into phase parameter $ \varphi $, is more efficient in single-qubit teleportation than the two-qubit one (see Fig. \ref{F11}). Therefore, depending on what parameter we want to teleport, we use single or two-qubit state to encode the required information. 

\par Finally, fidelity for the two-qubit teleportation under the Unruh channel, are found to be   
\begin{equation}
f=\frac{1}{N_{2}^{2}}\left[ \left(\frac{\mathcal{A}^{2}-\mathcal{D}^{2}}{4}+\mathcal{F}^{2}\text{cos}~2\varphi\right)2\text{sin}^{2}\theta  +\mathcal{D}^{2}\right]. 
\end{equation}
We get the results similar to single-qubit teleportation fidelity, investigating the behavior of two-qubit teleportation fidelity under the Unruh noise channel with and without applying the PMs.

\begin{figure}[ht!]
	\centering
	\subfigure{\includegraphics[width=6.5cm]{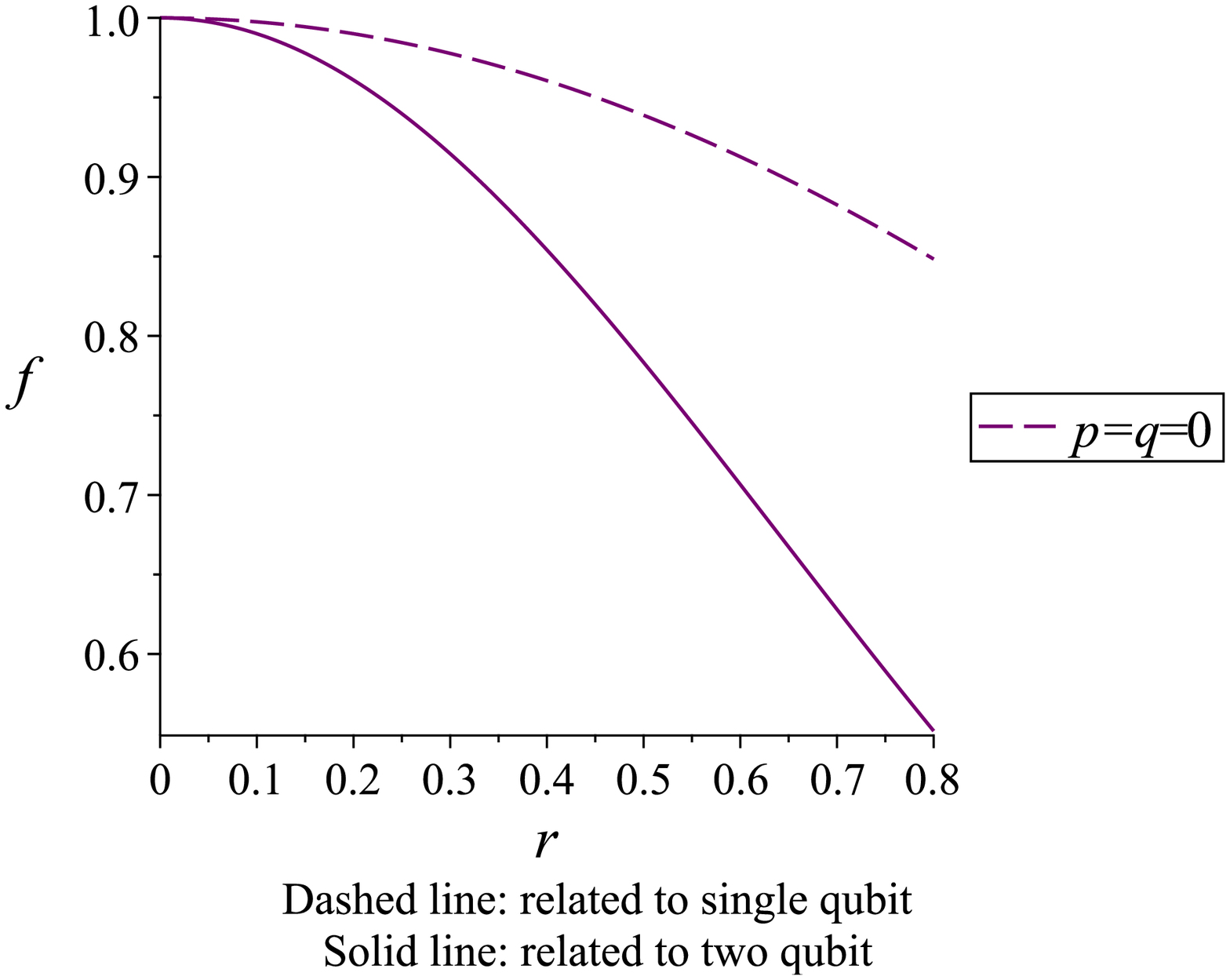}\label{fid1} }
	\subfigure{\includegraphics[width=6.5cm]{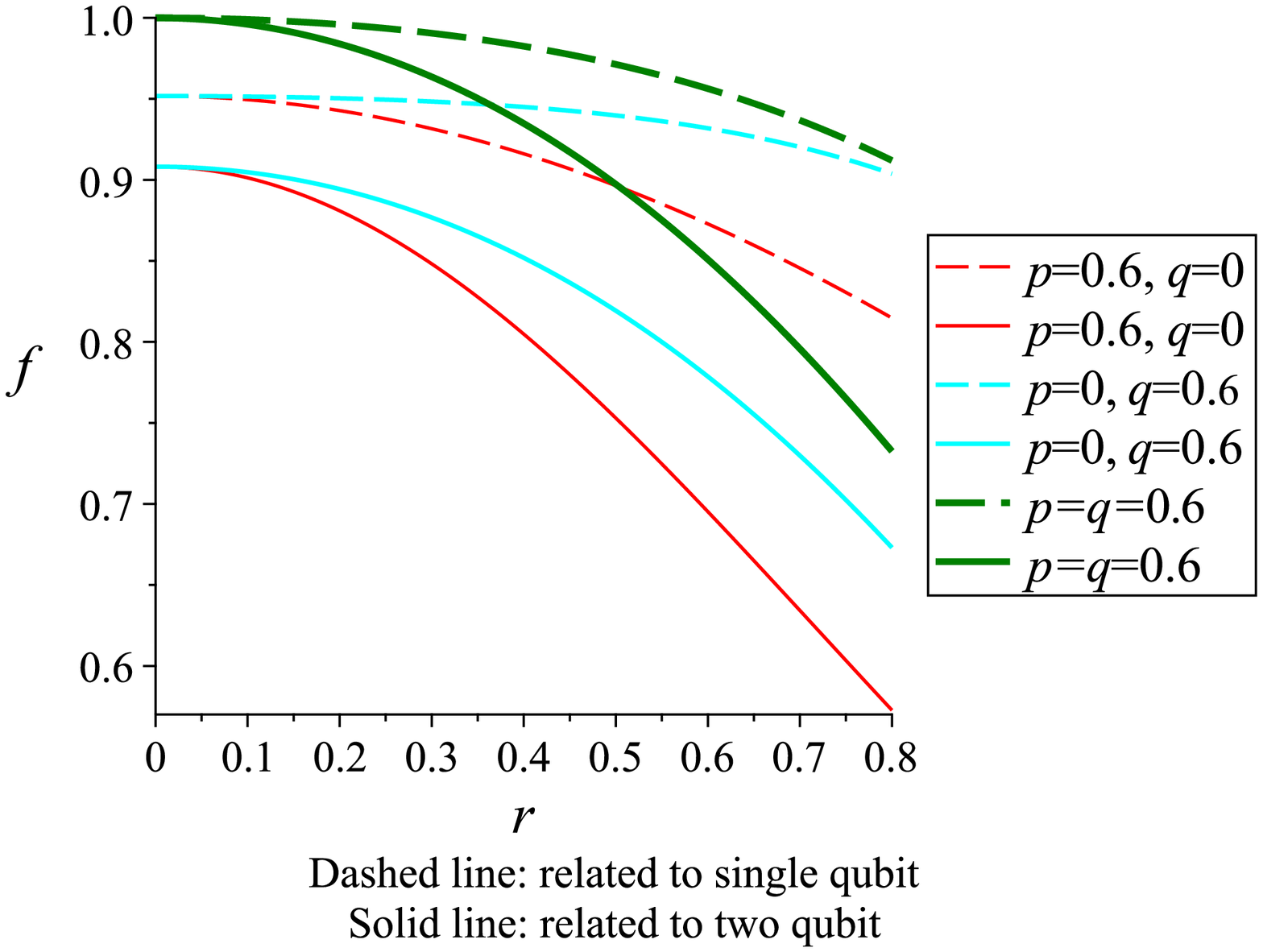}\label{fid2} }
	\caption{\small Comparing the fidelity of single- and two-qubit teleportation, fixing $ \theta=\frac{\pi}{2},\varphi=0 $ and $ \vartheta=\frac{\pi}{2}$ for (a) in the absence of measurements, (b) in the presence of measurements.}
	\label{fid}   
\end{figure}
\par Comparing the fidelity of single- and two-qubit teleportation in Fig. \ref{fid}, we observe that quality of teleportation is better in single-qubit case than the two-qubit one. It means that single-qubit teleportation is more robust against the Unruh decoherence.

\section{Summary and conclusions \label{conclusion}}
\par QRs and QFI of the teleported single- and two-qubit states, under the Unruh effect experienced by a mode of a free Dirac field, was discussed in this paper. We investigated the conditions under which the degradation effect of the Unruh effect on QRs and QFI of the teleported state can be improved by PMs, and found that the value of initial parameter of the channel $ \vartheta $ plays a key role in this scenario.
Moreover, we examined how the PMs
can be performed to eliminate  the Unruh effect or how they may be designed such that the Unruh effect can be used to enhance the quantum communication.
Besides, fixing the  acceleration and considering the behavior of the QFI, QC and teleportation fidelity as functions of PM strength (PMR strength), we found that $ F_{\text{out}}^{\text{PM}}\left(\varphi \right) $, QC and teleportation fidelity harmonically increase to reach a maximum value and then decrease with more increase in PM (PMR) strength. We also analytically analysed the optimal behavior of the 
QFI with respect to the phase parameter. Finally, comparing the QFI of the teleported single- and two-qubit states as functions of acceleration, we showed that the information encoded in the weight parameter $ \theta $ is better protected against the Unruh effect in the case of two-qubit teleportation. However, in the case of single-qubit teleportation, encoding information in the phase parameter $ \varphi $ is more efficient. Therefore, we encode the information into either the weight or phase parameter, depending on whether the two or single-qubit scenario, respectively, is used for the teleportation.

\section*{Acknowledgement(s)}

We wish to acknowledge the financial support of Urmia University and Jahrom University.

\noindent\textbf{Funding statement. }
M.J. and M.A.-T. wish to acknowledge the financial support of Urmia University. H.R. acknowledges funding by the grant no. 2471666442HRJ of Jahrom University.

\end{document}